\documentclass[12pt,preprint]{aastex}
\shorttitle{Infrared and radio study of IRAS 19111+1048 region}
\shortauthors{Vig et al.}
\begin{document}


\title{
Infrared and radio study of star forming regions associated with IRAS
19111+1048 and IRAS 19110+1045
}

\author{S. Vig$^1$,
S.K. Ghosh$^1$,
V.K. Kulkarni$^2$,
D.K. Ojha$^1$,
R.P. Verma$^1$}
\affil{1. Tata Institute of Fundamental Research,
Mumbai (Bombay) 400 005, India \\
2. National Centre for Radio Astrophysics, Pune 411 007, India}
\email{sarita@tifr.res.in}

\begin{abstract}
A multiwavelength study of the star forming regions associated with
IRAS 19111+1048 and IRAS 19110+1045 has been carried out. These have been
simultaneously mapped in two far infrared bands at $\lambda_{eff} = 130$ and
200 $\micron$ with $\sim 1\arcmin$ angular resolution
using the TIFR 1-m balloon borne telescope. The radio emission from the ionised
gas of these regions has been imaged at 1280, 610 and 325 MHz using the Giant
Metrewave Radio Telescope, India. A total of 20 compact radio sources have
been detected from the high resolution radio map
of IRAS 19111+1048 at 1280 MHz (of which one is of non-thermal origin).
 Assuming these sources to represent exciting zero age main
sequence (ZAMS) stars, the
initial mass function [$\xi(m) \propto m^{-a}$] is
found to be quite steep, with $a=5.3\pm0.5$ for the mass range
$14<m/M_{\odot}<33$. The near infrared (NIR) source coincident with the
 IRAS peak is likely to be an embedded pre-main sequence star.
An attempt has been made to identify sources responsible
for ionising the gas using NIR sources from 2MASS.
The spectral types of the ZAMS stars inferred independently from the
radio and NIR measurements match very well for a good fraction of the
radio sources having NIR counterparts.
For the IRAS 19110+1045 region, 
seven radio sources have been detected of which two are associated
with deeply embedded 2MASS objects.
Self consistent
radiative transfer modelling aimed at extracting important physical and
geometrical details of the two IRAS sources has been carried out using the
above cluster of ZAMS stars as the central
exciting sources. The best fit models are in good agreement with the observed
spectral energy distributions. A uniform density
distribution of dust and gas is implied for both the sources.
 The extents of ionised gas, number of ZAMS stars,
 presence of deeply embedded sources and lower value of $L/M$ for the
 cloud, support the youth of IRAS 19110+1045 vis-a-vis its neighbour, IRAS
19111+1048, consistent with earlier studies.
\end{abstract}

\keywords{ infrared: ISM -- radio continuum: ISM -- ISM: H II regions --
 ISM: individual (IRAS 19111+1048, IRAS 19110+1045)}


\section{Introduction}

Massive stars are born deeply embedded in dense clouds of gas and dust.
An outflow phase begins during early stage of stellar evolution.
However, outflows have been more commonly seen among low
mass stars than their massive counterparts (Shepherd 2005, and references
therein). This is because the evolutionary time-scales of massive
stars are much shorter. IRAS 19111+1048
(G45.12+0.13) and IRAS 19110+1045 (G45.07+0.13) are
massive star forming regions from where bipolar molecular outflows have been
detected (Hunter et al., 1997).

The H II region IRAS 19111+1048 is well-studied at many wavelengths as
compared to the neighbouring H II region IRAS 19110+1045. Both these sources
are at similar distances and a value of 6 kpc is assumed for the present work
(Simon et al., 2001; Araya et al., 2002; Fish et al., 2003).
According to Matthews et al. (1977), both the IRAS sources belong to
the same star forming complex and are separated by $\sim 3\arcmin.1$.
Kraemer et al. (2003) used the MIRAC3
instrument to obtain high angular resolution ($\sim$ 1$\arcsec$) images of the
ultracompact sources associated with IRAS 19111+1048 and IRAS 19110+1045 in
mid infrared (12.5 and 20.6 $\micron$) as part of the
 Galactic Ring Survey (GRS) and resolved the IRAS 19110+1045 core into
three compact sources and IRAS 19111+1048 core into two sources. Faison et al.
(1998) have presented the infrared spectroscopy ($3-13$ $\micron$) of these
regions and have studied the silicate and polycyclic aromatic
hydrocarbon (PAH) features as well as the [Ne II] line. They have also
discussed dust shell models to explain the spectral energy distributions (SED).

Low resolution CO observations ($\sim6\arcmin$ by Israel, 1982;
$\sim8\arcmin$ by Sanders et al., 1986) indicate that IRAS 19111+1048 is
located at the extremities of a large, clumpy molecular cloud complex
with IRAS 19111+1048 lying near the center of a well defined clump
(Hoare et al, 1991). With higher angular resolution ($\sim20\arcsec$)
CO observations (using the J$=3\rightarrow 2$ transition) Hunter et
al. (1997) found that there are several spokelike  protrusions from IRAS
19111+1048 core unlike the IRAS 19110+1045 core.

Wood \& Churchwell (1989) observed a cometary shape for IRAS 19111+1048 and a
compact unresolved source for IRAS 19110+1045 from their VLA survey at 5 and 15
GHz. High resolution ($\sim6\arcsec$ at 4.9 GHz \& $\sim3\arcsec$ at
 8.5 GHz) radio maps of these sources have been presented by Testi, Felli and
Taylor, 1999 (hereafter TFT). From these maps, they have
extracted four compact sources from the IRAS 19111+1048 region and two from
the IRAS 19110+1045. Extended radio emission has been seen around both
these regions. Both regions contain Type I OH masers (Goss et al., 1973) but
only IRAS 19110+1045 contains H$_2$O maser (Genzel \& Downes, 1977).
Similarly, it was found that methyl cyanide was detected only towards IRAS
19110+1045 and not towards the other source (Pankonin et al., 2001).
The presence of massive, bipolar molecular (CO) outflows from both these
sources was discovered by Hunter et al. (1997). Based on lack of H$_2$O and
methyl cyanide masers near the IRAS 19111+1048 core as compared to IRAS
19110+1045 as well as multiple outflows and higher CO antenna temperatures, it
is believed that of the two, IRAS 19111+1048 is a more advanced site of
massive star formation.

In the present paper, we carry out a comparative study of these two regions
using the infrared and radio observations. A multiwavelength
 study of these regions has been undertaken to gain further insight
into the evolutionary states of these two regions. This is accomplished using
the distribution of dust and ionised gas, dust optical depths, dust
temperatures and spectral types of zero age main sequence (ZAMS) stars in
these regions. While the dust around these regions is studied using
the mid and far-infrared emission, the ionised gas in these regions is probed
using the low frequency radio wavebands (1280, 610 and 325 MHz). The ZAMS
spectral types are determined from the radio observations and near infrared
(NIR) colour magnitude diagrams. Based on the spectral types determined from
the radio observations, we estimate the slope of the initial mass function
(IMF) for the IRAS 19111+1048 region.

In Section 2, we present the observations and other available data used for
the present study. Section 3 describes the observational results.
In Section 4, we present the radiative transfer models for the two
IRAS sources.
A comprehensive discussion of the multiwavelength study of these two regions
is carried out in Section 5. Section 6 summarizes our results.

\section{Observations and data reduction}

\subsection{Far infrared observations}
The Galactic star forming region associated with IRAS 19111+1048 has
been observed using two-band FIR photometer system at the Cassegrain focus of
the TIFR 100 cm (f/8) balloon borne telescope. The observations were carried
out during the balloon flight from the TIFR Balloon Facility, Hyderabad in
India (latitude $17^{\circ}.47$ north, longitude $78^{\circ}.57$ east) on March 08, 1998.
Details of the telescope and the observational procedure are given by Ghosh et
 al. (1988). A pair of 2$\times$3 composite silicon bolometer arrays, cooled
to 0.3 K by liquid $^3$He, corresponding to the two FIR bands are used to view
identical parts of the sky simultaneously. The field of view of each detector
is $1\arcmin.6$. The in-flight parameters of the telescope optics, necessary
to establish point spread functions etc, were determined from an
optical detector array (Photodiode array) in the Cassegrain focal
plane of the telescope. Details of the optical instrument can be found
elsewhere (Naik et al., 2000). The spectral response of each of the bands
 of the FIR photometer was determined in the laboratory using a Michelson
interferometer and a Golay cell as a comparison detector. Hereafter, the two
TIFR bands will be referred to as 130 and 200 $\micron$
bands corresponding to the $\lambda_{eff}$ for a modified blackbody source of
temperature 25 K and $\lambda^{-2}$ emissivity law.

The regions around IRAS 19111+1048 and IRAS 19110+1045 were mapped by raster
scanning the region of the sky ($\sim 30\arcmin \times 26\arcmin$) in
cross-elevation with steps in elevation at the end of each scan.
The sky chopped FIR signals were gridded into a matrix with pixel size
$0\arcmin.3 \times0\arcmin.3$.
The deconvolution of the observed signal matrix is carried out using
the maximum entropy method similar to that of Gull \& Daniell (1978) (for
details see Ghosh et al., 1988).
Absolute positional accuracy of $\sim 0\arcmin.8$ (Naik et al., 2000)
and angular resolution of $\sim1\arcmin$ have been achieved in the FIR
maps using this method.

\subsection{Radio observations}
The radio continuum observations in three frequency bands, 1280, 610 and 325
MHz were carried out using the Giant Metrewave Radio Telescope (GMRT), India.
The GMRT has a ``Y" shaped hybrid configuration of 30 antennas, each of 45 m
diameter. Six antennas are placed along the three arms (east, west and
south) and twelve antennas are located in a random but compact $1\times1$
km$^2$ pattern at the centre (Swarup et al., 1991). The baselines ($\sim$ 100 m
 - 25 km) provide sensitivity to large scale diffuse emission as well as high
angular resolution. The observational details are listed in Table 1.

For the observations, the primary flux density calibrators used were 3C48 and
3C286.
Since IRAS 19111+1048 and IRAS 19110+1045 are separated by a distance of
$3\arcmin.1$, both lie within the primary beam of the telescope at all the
frequencies. Hence, interferometric mapping with one phase center has been
carried out for both the sources. NRAO Astronomical Image Processing System
(AIPS) was used for data reduction. The data were carefully checked for
RF interference or other problems and suitably edited. The
calibrated data are Fourier transformed and deconvolved using the IMAGR task
in AIPS. Self calibration was carried out to remove the residual effects of
atmospheric and ionospheric phase corruptions and obtain the improved maps.
For the lower frequency (325 and 610 MHz) images, the system
temperature was obtained using the sky temperature from 408 MHz map of Haslam
et al. (1982). A correction factor equal to ratio of the system temperature
towards the source has been used to scale the deconvolved images. For the
sources detected (S/N $>10$) in the radio
maps, the peak position as well as the flux densities (peak and integrated)
were obtained using the task JMFITS in AIPS.

\subsection{Other available datasets}
Other available data (2MASS, IRAS, MSX, ISO, JCMT) for the regions around
IRAS 19111+1048 and IRAS 19110+1045 have been used to study them at different
wavelengths ($1.2-850$ $\micron$). The flux densities from the images have
been used for the construction of the SED in
order to extract physical parameters of the star forming regions using
radiative transfer modelling (see Section 4). The 2MASS point sources and images
have been used to identify the sources of ionisation of gas in these regions.

\subsubsection{IRAS}
The data from the Infrared Astronomical Satellite (IRAS) survey in the four bands (12, 25, 60 and 100 $\micron$)
for IRAS 19111+1048 and IRAS 19110+1045 were HIRES (High Resolution
processing using Maximum Correlation Method; Aumann
et al., 1990) processed at the Infrared Processing and Analysis Center
(IPAC, Caltech) to obtain high angular resolution maps
($0\arcmin.5\times1\arcmin.0$ at 12 \micron, $0\arcmin.5\times1\arcmin.1$ at 25 \micron, $1\arcmin.0\times1\arcmin.6$
 at 60 $\micron$, $1\arcmin.9\times2\arcmin.3$ at 100 $\micron$). Apart from
extracting the sources and determining the flux densities, the IRAS-HIRES maps
at 12 and 25 $\micron$ have been used to find the
spatial distribution of temperature, T(12/25), and optical depth at 25
$\micron$ ($\tau_{25}$).

IRAS 19111+1048 appears in IRAS Low Resolution Spectrometer (LRS) Catalog
(IRAS Science Team, 1986)
and the  spectrum of this source in the wavelength range
$8-22$ $\micron$ has been used for constructing the SED.

\subsubsection{MSX}

The Midcourse Space Experiment\footnote{ This research made use of data
products from the Midcourse
Space Experiment. Processing of the data was funded by the Ballistic
Missile Defense Organization with additional support from NASA Office of Space
Science. This research has also made use of the NASA/ IPAC Infrared Science
Archive, which is operated by the Jet Propulsion Laboratory, Caltech, under
contract with the NASA.} (MSX) surveyed the entire Galactic plane
within $|$b$|\le 5^{\circ}$ in four mid infrared wavebands: 8.3, 12.1, 14.7
and 21.3 $\micron$ at a spatial resolution of $\sim18\arcsec.3$ (Price et al.,
2001). The panoramic images of the Galactic plane survey of MSX were taken
from IPAC ({\tt http://irsa.ipac.caltech.edu/applications/MSX/}). These MSX
maps were used to extract the sources and determine their flux densities for
constructing the SED. Point sources close to these star
forming regions have been selected from MSX Point Source Catalog Version 2.3
(Egan et al., 2003) and cross-correlated with 2MASS sources.

\subsubsection{2MASS}
The point sources around the regions IRAS 19111+1048 and IRAS 19110+1045 were
extracted from the Two Micron All Sky Survey\footnote{This publication makes use of data products from the Two Micron All Sky
Survey, which is a joint project of the University of Massachusetts and the
Infrared Processing and Analysis Center/California Institute of Technology,
funded by the NASA and the NSF.} (2MASS) Point Source Catalog
(PSC).
The 2MASS PSC is complete down to J $\le15.8$, H $\le15.1$ and K$_s\le14.3$
mag for S/N$>10$, in the absence of confusion. The J, H and
K$_s$ magnitudes of the extracted sources have been used to make
colour-magnitude diagrams which, in conjunction with the radio observations,
are used
to identify the sources responsible for ionisation of gas in these regions. The
JHK$_s$ magnitudes and images were taken from IPAC
({\tt http://www.ipac.caltech.edu/2mass/}).

\subsubsection{ISO}

For IRAS 19110+1045, the data from the Short Wavelength
Spectrometer (de Graauw et al., 1996) and the Long Wavelength Spectrometer
(Clegg et al., 1996) of the Infrared Space Observatory\footnote{Based on
observations with ISO, an ESA project with instruments funded by ESA Member
States (especially the PI countries: France, Germany, the Netherlands and the
United Kingdom) and with the participation of ISAS and NASA.} (ISO)
between $2-188$ $\micron$ have been used for constructing the SED of this
source.
The
version of the ISO data used in this paper correspond to the Highly Processed
Data Product (HPDP) sets called `A uniform database of SWS 2.4-45.4 micron
spectra', `High resolution processed and defringed SWS01s' and `Uniformly
processed LWS L01 spectra' by Sloan et al. (2003), Frieswijk et al. (2004) and
Lloyd et al. (2003) respectively, obtained from the ISO Data Archive.

\subsubsection{JCMT}

Submillimeter observations using the Submillimetre Common-User Bolometer Array
(SCUBA) instrument of James Clark Maxwell Telescope\footnote{This paper
makes use of data from the James Clarke Maxwell Telescope (JCMT) Archive. The
JCMT is operated by the Joint Astronomy Centre on behalf of the UK Particle
Physics and Astronomy Research Council, the National Research Council of
Canada and the Netherlands Organisation for Pure Research.}
(JCMT) of IRAS 19111+1048
 and IRAS 19110+1045 were obtained from the JCMT archive and processed using
their standard pipeline SCUBA User Reduction Facility (SURF). Maps generated
at 450 and 850 $\micron$ were used to extract the sources and determine
their flux densities for constructing the SEDs.

\section{Results}

\subsection{Emission from cold dust}

The deconvolved TIFR far-infrared maps at 130 and 200 $\micron$ bands are
presented in Fig 1. The sources corresponding to IRAS 19111+1048 and IRAS
19110+1045 are resolved in the 130 $\micron$ map and an extension
corresponding to IRAS 19110+1045 can be seen in the 200 $\micron$ map. The
flux densities of both the sources at 130 $\mu$m, and of IRAS 19111+1048 at
200 $\mu$m
obtained by integrating over circular regions of diameter
$3\arcmin$, centered on the far infrared peaks,
are listed in Table 2. These far infrared wavebands probe the
distribution of cold dust around this region. The colour temperature of dust
is obtained as $\sim 34$ K for this region ($\sim3\arcmin$) using a dust
emissivity law, $\epsilon_{\lambda} \propto \lambda^{-2}$ and assuming that
the dust is optically thin.

The JCMT maps of these sources at 450 and 850 $\micron$ detect the cold dust
around these regions. The distribution of the emission from dust at these
wavelengths is shown in Fig 2. The
maps at 450 and 850 $\micron$ correspond to beam sizes of $\sim7''\times11''$
and $\sim13''\times17''$, respectively.

\subsection{Emission from warm dust}

The HIRES processed maps of IRAS 19111+1048 and IRAS 19110+1045 at 12, 25, 60
and 100 $\mu$m are shown in Fig 3, where both the sources are clearly
resolved. Table 2 lists the flux densities determined from the HIRES maps. The
IRAS PSC flux densities of these sources in all four wavebands are also
presented for comparison.

We have used the IRAS-HIRES maps at 12 and 25 $\mu$m to generate the spatial
distributions of dust colour temperature, $T(12/25)$ and optical depth at 25
$\mu$m ($\tau_{25}$). The emissivity law of $\epsilon_{\lambda} \propto
\lambda^{-1}$ was assumed to generate these maps. It may be noted that the
spatial distribution of temperature is insensitive to the assumed value of the
emissivity exponent. These wavelengths probe the warm ($100-160$
 K) dust in the region around IRAS 19111+1048 and IRAS 19110+1045. The peak
optical depth at 25 $\mu$m is $0.001$ located at the peak infrared
emission of IRAS 19111+1048. We have presented the distribution of $\tau_{25}$
as contours in Fig 4.

The emission in mid infrared from warm dust, in the MSX 8.3 $\mu$m band,
towards these sources is shown in Fig 5 (left). The flux densities from the
four MSX maps integrated for circular regions with $3\arcmin$ diameter around
IRAS 19111+1048 and IRAS 19110+1045 are listed in Table 2.
We have modelled the thermal continuum from the interstellar dust alongwith
the emission in the Unidentified Infrared Bands (UIBs), using the data from MSX
 in the 8, 12, 14 and 21 $\micron$ bands, following the scheme
developed by Ghosh \& Ojha (2002). The peak strength of the modelled UIB
emission is $5.8\times10^{-4}$ W m$^{-2}$ Sr$^{-1}$ at the position of IRAS
19111+1048. The spatial distribution of UIB emission resembles that of the
continuum emission from warm dust at 8.3 $\mu$m. Faison et al. (1998)
carried
out infrared spectroscopy of central regions ($\sim 9\arcsec$) of both
these sources but have not detected the PAH features. The peak PAH emission
obtained from MSX is $\sim5$ times the noise level of Faison et al. (1998). As
the resolution of MSX is
$\sim18\arcsec$, it is likely that the UIB/PAH emission has a shell
type distribution as seen in other star forming regions (Darbon et al. 2000).
 The temperature of the mid infrared emitting dust is found to be in the range
100-160 K (consistent with that obtained from IRAS-HIRES T(12/25)), the
spatial distribution of which is presented as isotherms in Fig 5 (right).

\subsection{Emission from ionised gas}

The radio continuum emission from the region around IRAS 19111+1048 and IRAS
19110+1045 at 1280 MHz is shown in Figs 6 and 7, respectively. The
corresponding maps at 610 and 325 MHz are shown in Figs 8 and 9,
respectively. The synthesized beam sizes and the rms noise in the maps are
given in Table 1. As can be seen from the radio emission maps at 1280 MHz, 20
discrete sources have been detected in the region around IRAS 19111+1048 and 7
in the IRAS 19110+1045 region (with S/N$>10$). A list of these sources
along with their peak and integrated flux densities at 1280 and 610 MHz are
presented in Tables 3 and 4. We designate these sources as S1, S2,
... , S27.
The total integrated flux densities for IRAS 19111+1048 at 1280, 610 and 325
MHz are $6.60\pm0.01$, $3.90\pm0.01$ and $0.8\pm0.02$ Jy, respectively.
The extent of ionised gas of the H II region IRAS 19111+1048 is $\sim2.8$ pc
while that of IRAS 19110+1045 is $\sim1.3$ pc.
The flux density measurement of IRAS 19111+1048 was carried out by
Altenhoff et al. (1979) at 4.875 GHz with the Effelsberg 100-m telescope (beam
$\sim2'.6$). Scaling this measurement to 1.280 GHz
under the assumption that it is optically thin at both these frequencies, we
find that with the interferometer, we have collected $\sim95\%$ of the total
flux density as measured with the single dish demonstrating that almost all
 structural features have been captured by the baselines of GMRT.
For IRAS 19110+1045, the total integrated flux densities are $730\pm6$ mJy,
$552\pm6$ mJy \& $110\pm4.3$ mJy at 1280, 610 \& 325 MHz, respectively.

Assuming that each discrete source in the radio continuum maps represents an
exciting ZAMS star, the spectral types of the sources have
been estimated using the integrated flux densities from our GMRT maps and the
formulation of Schraml \& Mezger (1969) and Panagia (1973). The ZAMS spectral
types determined in this way are listed in Tables 5 and 6. As can be
seen all the
sources are earlier than B0.5. For the sources also detected by TFT, we
have compared the spectral types obtained from our GMRT maps as well as
from the flux densities given by TFT and these are found to be consistent
(see Tables 5 and 6).

\subsubsection{Physical properties of the compact cores}

The physical properties like the electron temperature and emission measure of
the compact core of IRAS 19111+1048 (shown as `14' in Fig 6 and S14 in Table 3)
 can be
estimated using the flux densities of the unresolved core at low frequencies
(1280 and 610 MHz from the GMRT images) as well as at higher frequencies
(4.9 and 8.5 GHz using the listed parameters of this core from Table 1 of TFT).
 Under the approximation that the core is homogeneous and spherically
symmetric, the flux density, S, (Mezger \& Henderson, 1967) is given by
$$S = 3.07\times10^{-2}T_e\nu^2\Omega(1-e^{-\tau(\nu)})$$
$$\tau(\nu)=1.643\times10^5\, a\, \nu^{-2.1}\:(EM)\:T_e^{-1.35}$$
where $S$ is the integrated flux density in Jy, $T_e$ is electron temperature
in K, $\nu$ is the frequency in MHz, $\tau$ is the optical depth, $\Omega$ is
the solid angle subtended by the source in Steradian (which in this case is the
synthesized beam size for the unresolved core), and $EM$ is the emission
measure in cm$^{-6}$pc. The factor $a$ represents the deviation
between the exact
formulation and its approximation. We use $a=0.99$ obtained from Table 6 of
Mezger \& Henderson (1967) for the frequency range $0.6-8$ GHz for
$T=10,000$ K. The peak flux densities of the cores after convolving the
images to a common resolution ($10\arcsec\times 10\arcsec$) have been used to
fit the above equations using nonlinear least-squares Marquardt-Levenberg
algorithm. The observed and the best fit model flux densities are shown in
Fig 10. For the best fit model, the electron temperature and emission measure
(with 1$\sigma$ errors) of the IRAS 19111+1048 core are $8346\pm1511$ K and 
$8.8\pm0.7\times10^6$
cm$^{-6}$pc, respectively. Wood \& Churchwell (1989) used the H76$\alpha$
recombination line to obtain the electron temperature of $7900\pm1000$ K which
is consistent with our value within errors. Using radio
continuum emission at 15 GHz, they obtained the peak
($0\arcsec.4\times0\arcsec.4$) values of
electron temperature and emission measure to be 8100 K and
$1.5\times10^9$ cm$^{-6}$pc, respectively. The difference in $EM$
values could be due to clumpy distribution of ionised gas.
Lumsden \& Puxley (1996) obtained an
electron temperature of $8000-9000$ K by means of a best fit to the near
infrared spectrum of IRAS 19111+1048 core.

The compact cores S26 and S27 belonging to IRAS 19110+1045 region
are resolved at 1280 MHz but not at 610 MHz. They are, however, resolved
in the 4.9 and 8.5 GHz maps of TFT. While S26 appears to be optically thin at
these three frequencies (1280 MHz, 4.9 GHz and 8.5 GHz), S27 appears to be
optically thick. For an assumed temperature of 8000 K, the best fit
emission measure for S26 is $2.5\pm0.3\times10^5$ cm$^{-6}$pc. Since S27 is a
compact unresolved source at all these three frequencies, we estimate the size
of S27 to be 0.026 pc using the flux density at 1280 MHz for a similar
temperature. Assuming photon bounded uniform density region, this size implies
 electron density, $n_e=1.5\times10^4$ cm$^{-3}$. Garay et al. (1986) modelled
this source as an ionised toroid of size $0.03$ pc (scaled to our distance)
and obtained $n_e=1.6\times10^5$ cm$^{-3}$, which are comparable.

\subsection{Embedded star cluster}

With the aim of identifying early type stars in the embedded cluster of the two
H II regions, we study the NIR point sources from 2MASS PSC in the J,
H and K$_s$ bands. In this section, we discuss the nature of these sources. We
have used the colour-magnitude (CM) diagrams of sources in the regions around
IRAS 19111+1048 and IRAS 19110+1045 to estimate the spectral types of the
stars. For IRAS 19111+1048, 47 stars were found in a
circle of radius $1\arcmin.2$ around the center of the radio emission
($\alpha_{2000}$ = $19^h$ $13^m$ $26.0^s$, $\delta_{2000}$ = +10$^{\circ}$
54$\arcmin$ 11$\arcsec$) which were detected in all the three JHK$_s$ bands of
2MASS. All the sources whose magnitudes are given as upper limits have
been excluded from the study. A similar study was carried out for the source
IRAS 19110+1045. Within a circle of radius $1 \arcmin$ around this source
($\alpha_{2000}$ = $19^h$ $13^m$ $20.6^s$, $\delta_{2000}$ = +10$^{\circ}$
50$\arcmin$ 47$\arcsec$), 42 stars were detected in all the three 2MASS
bands.
A list of the positions and apparent magnitudes of the 2MASS sources
that are likely to be associated with the radio sources of IRAS 19111+1048 and
IRAS 19110+1045 regions (see Section 5) is presented in Table 7. The CM
diagrams (J-H vs J) for the sources
detected in 2MASS JHK$_s$ bands around IRAS 19111+1048 and IRAS 19110+1045
regions are shown in Fig 11. The nearly vertical solid lines from left to
right represent the ZAMS curves (for a
distance of 6 kpc) reddened by A$_V$ = 0, 15 and 30 mag, respectively. The
slanting lines trace the reddening vectors of these zero main
sequence stars. We have assumed extinction values of A$_J$/A$_V$ = 0.282,
A$_H$/A$_V$ = 0.175 and A$_{K_s}$/A$_V$ = 0.112 from Rieke \& Lebofsky (1985).
All the 2MASS magnitudes as well as the ZAMS curves are in the Bessel \& Brett
(1988) system. In Fig 11, the triangles (IR4, IR8) represent sources which are
very bright in K$_s$ band and lie much higher than the extinction curve of O5
star, possibly embedded or unresolved stellar objects. The asterisk symbols
represent sources of spectral type B0 or earlier and lying within
the radio nebulosity. The plus symbols represent stars
of spectral type later than B0 but found within the radio nebulosity. The
dot symbols represent sources which do not lie within the radio nebulosity.

\section{Radiative transfer modelling}

With the aim of extracting important physical parameters, an attempt has been
 made to construct radiative transfer models of the two sources IRAS
19111+1048 and IRAS 19110+1045. Although high resolution maps in radio and
NIR indicate the presence and location of a cluster of sources, we
carry out simplistic radiative transfer modelling assuming a cluster of ZAMS
stars located at the center of a spherically symmetric homogeneous cloud of
gas and dust. This cloud is assumed to be immersed in an isotropic
 radiation field (typical Interstellar Radiation Field, ISRF). The
self-consistent scheme developed by Mookerjea and Ghosh, (1999) has been used
here.

The physical parameters explored to get a fit to the SED are: the nature of the embedded source (single or a cluster of
ZAMS stars of appropriate initial mass function and mass limits), radial
density distribution law [$n(r) \propto r^{-\alpha}$,
$\alpha = 0$, 1, 2], relative abundances of the two constituent grain types
(silicate and graphite), the total radial optical depth due to the dust, the
gas-to-dust ratio by mass and geometric details of the cloud (e.g. cavity
size, outer size of the cloud).  Two commonly used types of interstellar dust
are explored: the type of dust grains from Draine \& Lee, (1984; DL)
and the dust type from Mathis, Mezger \& Panagia, (1983). The observed
angular sizes have been used to constrain the models.

\subsection{IRAS 19111+1048}

The SED for IRAS 19111+1048 has been constructed using the flux densities
obtained from the MSX maps, the IRAS-LRS data, the IRAS-HIRES maps,
the TIFR maps, the submillimeter (450 and 850 $\mu$m) JCMT-SCUBA maps.
The SED also includes the flux densities in the J, H and K$_s$ bands
obtained by integrating the
fluxes of the 2MASS sources within a circle of diameter $3\arcmin$ centered on
the peak of IRAS 19111+1048.
The total luminosity obtained by integrating the observed SED is
$6.5\times10^5 L_{\odot}$. We have used a cluster of 19 ZAMS
OB stars, obtained from our radio measurements (first 19 sources listed in
Tables 3 and 5, S20 being a non-thermal source; see Section 5 for more
details), as the centrally exciting sources which power this region.
The total luminosity of this cluster is $8.5\times10^5 L_{\odot}$. Modelling
the SED of IRAS 19111+1048 gives a better fit for uniform ($r^0$) density
distribution than for the other two power laws. The outer
radius of the spherical cloud is 4 pc and and radius of the dust cavity
$\sim0.01$ pc. The preferred dust is
of DL type with radial optical depth of 0.002 at 100 $\micron$. We have used
the radius of the H II region (1.4 pc) from our 1280 MHz map to constrain the
gas-to-dust ratio by mass, which is obtained to be 250. This
implies an under-abundance of dust grains in this region.
The radio continuum emission predicted by this model at 1280 MHz is
3.8 Jy which is $\sim60$ \% of the observed value. The best fit model along
with the SED is shown in Fig 12.
The predicted emission from the model at wavelengths shorter than
$\sim2$ $\micron$ is dominated by the scattered ISRF due to dust grains in the
cloud. IRAS 19111+1048 is unresolved at the MSX and
IRAS-HIRES wavebands; this is consistent with the sizes obtained from the
model. The expected angular sizes of IRAS 19111+1048 in the TIFR maps have
been estimated from the model by convolving the predicted source size at that
band with the achieved angular resolution. This is consistent with the sizes
seen at 130 and 210 $\micron$. The various parameters of the best-fit model
are listed in Table 8.

\subsection{IRAS 19110+1045}

For IRAS 19110+1045, the ISO-SWS and ISO-LWS spectra from 3 to 188 $\mu$m are
available.
The radiative transfer modelling uses the flux densities obtained
from the JCMT-SCUBA maps at 450 and 850 $\micron$ bands to carry out radiative
transfer modelling for this source.
The SED also includes the total emission from 2MASS
sources within the equivalent beam of ISO-SWS, in the J, H, and K$_S$
bands. From our radio maps, seven radio sources have been extracted from this 
region (see Table 6). In our model, we have used these as the central exciting 
sources. ISO-SWS spectrum covers the wavelength range $2-45$ $\micron$
corresponding to a field of view (FOV) $33\arcsec \times 20\arcsec $. On the
other hand, the ISO-LWS spectrum covers wavelengths longer than $\sim 45$
 $\micron$ with a FOV of diameter 84$\arcsec$.
For comparing the radiative transfer model results with the LWS and SWS
data, the model results have been convolved with the appropriate beams.
Consequently, we see a slight discontinuity in the SED at $\sim 45$ $\mu$m.
We obtain the total luminosity to be $3.3\times 10^5 L_{\odot}$ by integrating
the observed SED. However the total luminosity of the resolved
stars is $1.1\times 10^5 L_{\odot}$.
Guided by the composition of the observed
radio spectral types as well as the luminosity obtained by integrating the SED,
 we carry out radiative transfer modelling. The data and the best fit model
are shown in Fig 13. The model fits the observations reasonably well. The
model fitted to the observed SED best corresponds to a constant density cloud.
The radio continuum flux density predicted by the model underestimates
the observed flux density by a factor of $3$ at 1280 MHz but matches
the flux density at 8.3 GHz (TFT) of 0.84 Jy for a gas-to-dust ratio of
$450$.
This gas-to-dust ratio is rather large. The
radius of the H II region predicted by this model is 0.04 pc while the radius
of the region encompassing S21-S25 is found to be $\sim0.9-1.1$
pc at 1280 MHz and 4.9 GHz. The large gas-to-dust ratio and the
difference in the size of the HII region from the model and observations
implies significant difference of the true geometry of the source from the
simple assumptions used in the modelling.
It is likely that the medium is
clumpy. The DL type of dust fits the data better. Table 8 lists the various
parameters of the best-fit model to the SED.

\section{Discussion}

\subsection{IRAS 19111+1048}

The far infrared IRAS-HIRES, TIFR and JCMT-SCUBA maps probe the cold dust
environment around IRAS 19111+1048.
The flux densities from the JCMT-SCUBA maps at 450 and 850 $\micron$
have been used to compute the mass of dust using the formulation of Hildebrand
(1983) and Sandell (2000). The dust mass obtained from emission at 450 and 850
$\micron$ is $\sim$ 20 $M_{\odot}$ while that obtained from 200
$\micron$ emission map is 18 $M_{\odot}$. A dust mass of 12 $M_{\odot}$ is
obtained from
the radiative transfer modelling. This compares well with the dust mass of
24 $M_{\odot}$ obtained by Hoare et al., 1991, (scaled to our distance) from
the submillimeter observations.
Considering a gas-to-dust ratio of 250 by mass from the radiative transfer
model (see Section 4.1), we determine the gas mass in this region to be
5,000 $M_{\odot}$.

The optical depth at 100 $\micron$ obtained from radiative transfer
modelling is, $\tau_{100}\sim 0.002$. This could be compared with the peak value of optical depth at 25
$\micron$ obtained from the IRAS-HIRES
maps, $\tau_{25}\sim 0.001$. The apparent inconsistency of the lower value of
$\tau_{25}$ as compared to $\tau_{100}$ can be resolved by attributing the
$\tau_{25}$ to warmer dust of smaller volume around the exciting sources.
Faison et al. (1998) modelled the dust distribution
around this source and obtained a value of $\tau_{100}=0.01$. This
discrepency can be attributed to (i) distance of 9.7 kpc used by them, and (ii)
 their model fits only the mid-infrared spectrum reasonably well and not the
full SED.

The radio emission of this H II region indicates the extent of ionised gas.
As can be seen from all the radio maps, a number of sources
have been detected apart from diffuse emission. A comparison with the high
resolution maps of TFT at 4.9 and 8.5 GHz
indicates that our source list includes all the four sources detected by them.
The compact sources detected by them around this UC H II region are listed as
sources 13, 15, 16 and 17 in their Table 1. We shall designate these sources as
TFT13, TFT15, TFT16 and TFT17, respectively. A positional comparison of their
sources and those detected by us at 1280 MHz indicates that TFT13 can be
identified with S6, TFT15 with S14, TFT16 with S18 and TFT17 with S19
(Table 5).
The diffuse emission corresponding to TFT31 is emission to the north-west of
S14 in our maps which has been resolved into multiple sources here.
A comparison between the 1280 and 610 MHz maps show that many sources which
have been detected in the former are not resolved in the latter.

For the point source S20, we find that the flux density at 610 MHz is higher
than the flux density at 1280 MHz band corresponding to a power law spectral
index of -0.6. This indicates non-thermal nature of the radio emission.
Assuming this spectral index, the expected flux densities at 4.9 and
8.5 GHz are 8.6 mJy and 6.2 mJy, respectively. The absence of this source from
the maps
of TFT is consistent with the lowest peak values of the compact sources listed
in their Table 1: 17 mJy/beam at 4.9 GHz and 24 mJy/beam at 8.5 GHz.
Though S20 is in
the vicinity of IRAS 19111+1048, it is unlikely to be associated with IRAS
19111+1048 and consequently, we shall not consider S20 as a part of
this H II region for further analysis. In the 325 MHz image, we do not see this
source due to resolution effect though an extension corresponding to this
source can be seen.

 Using the 19 radio sources, we find the slope of the initial mass
function [$\xi(m) \propto m^{-a}$] in this star forming region to be
$a=5.3\pm0.5$ in the mass range $14<m/M_{\odot}<33$. The initial mass function
is quite steep implying a scarcity of massive and luminous stars in
this region. Such a steep power law of the initial mass function is
characteristic of young star forming regions. For example, Ghosh et al. (1989)
obtained $a=6$ and 3 for the regions W31 and W33, respectively,
for $20< m/M_{\odot}<35$ using a similar analysis.

We now identify the NIR counterparts of these radio sources using
the position information ($<2\arcsec.5$; resolution of 1280 MHz map)
 from the 2MASS NIR images and catalogs. We designate these NIR sources as IR1, IR2, ... , IR7 (Table 7). The 1280 MHz radio contours are
plotted over the 2MASS K$_s$-band image, shown in Fig 14. Some of the radio
sources with their possible ionisation sources from NIR (position and colour)
are also compared in Table 5. Six sources have been identified of
which the spectral types deduced from radio emission of the three sources,
S6, S11 and S14, agree quite well with the spectral types obtained from NIR
measurements. We have also checked for
positional match of the radio sources with sources detected in H \& K$_s$
bands or only in K$_s$ band. We find only one such match for S4 which
is detected in only K$_s$ band (mag $\sim14.1$).

The compact H II source, S14, is deduced to be of spectral type O6 from the
integrated radio emission. There are two NIR objects, IR4 and IR5,
within S14 region as obtained from the NIR catalog and images. IR4
is at the peak of the radio emission and matches the OH maser position
($\alpha_{2000}$ = $19^h$ $13^m$ $27.8^s$, $\delta_{2000}$ = +10$^{\circ}$
53$\arcmin$ 37$\arcsec$) obtained by Argon et al. (2000). IR4 is very bright
in K$_s$-band (mag $\sim7.9$) and appears to be of
spectral type much earlier than O5 as indicated by a triangle in the CM
diagram. It could be an embedded pre-main sequence star or it is likely to
comprise of two or more unresolved early type stars. Hunter et al. (1997)
detected two sources of CO bipolar outflow; one of which is at the location of
the radio peak IR4. Considering the facts: (a) IR4 is coincident with the OH
maser position, (b) it is the source of bipolar outflow, and (c) has large
infrared excess (J-K $\sim4.0$), it is likely that this source could be an
embedded pre-main sequence object.
The ionisation of gas could also  be attributed to IR5,
which according to our CM diagram is of stellar type O5. It is interesting to
note that although the source S14 peaks at the position of IRAS source, the
radio morphology in all the three GMRT bands, shows a north-south extension
indicating that perhaps this is not a single compact core but rather has more
than one H II clump. This is also clearly evident from the maps of TFT.
Kraemer et al. (2003) find that the IRAS 19111+1048
core consists of two compact mid infrared (12.5 and 20.6 $\micron$) components.
 If we consider sources only detected in the K$_s$ band within the S14 region,
we find that there are four more NIR sources detected. It thus appears
that the source S14 has a compact embedded cluster and is perhaps a
conglomeration of H II clumps, the brightest one at the position of S14.

For the sources S6, S11 and S14, the spectral types obtained from radio
measurements and NIR colours agree very well. However, the agreement is not
good for S16, S19 and S12. S16 and S19 are found
to be of spectral types B0.5-B0 and B0-O9.5, respectively.
They are associated with IR6 \& IR7, respectively, which are of spectral type
B7-B3 obtained from the NIR CM diagram. Most probably, IR6 and IR7 are
reddened main sequence foreground
sources based on their position in the CM diagram. For S12 also, we find
that the radio and NIR spectral types do not match. While the radio
spectral type of the ZAMS star is O8, the NIR counterpart is  of spectral type
B2. These three cases could be spurious associations. This number of
spurious associations can be compared with the estimated number of chance
associations of 1.3 obtained by considering the density of NIR sources (detected in H and K$_s$ bands) and the search radius $\sim2\arcsec.5$ of the
19 radio sources.

From the NIR CM diagram, we see that most of the ZAMS stars within
the radio nebulosity have extinction values corresponding to
$A_V\sim10-35$ mag.
This is quite consistent with the extinction values obtained ($A_V = 10-30$)
by Hoare et al. (1991). Lumsden \& Puxley, (1996) obtained $A_V\sim24$
from Br-$\gamma$ imaging and radio continuum observations for this source.

Although the radio and NIR K$_s$-band images indicate positional match for
some radio and NIR sources (with consistent spectral types), there are a
number of radio sources which do not have a NIR counterpart. The radio
spectral types indicate that these
are bright sources which ought to have been detected in NIR. Their
absence gives rise to two possibilities: (1) the radio sources are high density
 ionised clumps with no centrally exciting star or (2)
the medium is clumpy with dense distributions of A$_V>30$, while the
average visual extinction is lower (A$_V\sim15$). Next, we show that the first
scenario of high density clumps being ionised by
ZAMS stars located away from them can be ruled out. For instance, S13 does
not have a NIR counterpart. An estimation of the Lyman continuum photons
reaching S13 from its neighbour S14 (the brightest source in this region; see
Fig 6) due to geometrical factors shows that, in the absence of any
intervening medium, the expected luminosity is only 5\% of its
observed luminosity. Hence, it is more likely that the
second scenario is responsible for the non-detection of the NIR counterparts.
A high resolution mapping of molecular species probing the cold and
dense gas over a large region ($\sim1.5\arcmin$)
 around this source is required to better our understanding of the physical
picture of this region. The radio morphology towards the north-east of the
extended radio emission indicates steepness of the gradient of the radio
emission. This could be due to the restriction of the ionised region by
molecular gas, as seen in other regions like S201 (Ojha et al., 2004a) and
NGC7538 (Ojha et al., 2004b).

We have also cross-correlated the MSX PSC sources with those from
2MASS PSC which lie within the radio nebulosity. There are 2 such MSX PSC
sources, G045.1331+00.1436 and G045.1221+00.1323 (hereafter M1 and
M2, respectively; listed in Table 9 and shown in Fig 14). While M1 is
close to IR2 ($\sim5\arcsec$),
M2 coincides with IR4. From the 2MASS-MSX colour-colour diagram
($F_{21}/F_8$ vs $F_8/F_K$, Lumsden et al. 2002), we find that these sources
lie in the region generally covered by compact H II regions. We can, therefore,
 conclude that these are young stars associated with the cluster.

\subsection{IRAS 19110+1045}

The SED of IRAS 19110+1045 shows a very strong silicate
feature at 10 $\micron$. From radiative transfer modelling, we get a higher
composition of silicate (60\%) grains. The optical depth at 100 $\micron$ from
the model is obtained to be 0.1. This is consistent with the value of
$\tau_{100}=0.085$ obtained by Faison et al. (1998) using their spherically
symmetric dust shell model.
A gas to dust ratio by mass of 450 is obtained for consistency with radio
observations. The dust mass corresponding to the
above model is 450 $M_{\odot}$.
On the other hand, the 130 $\micron$ emission map gives a dust mass of
13 $M_{\odot}$. However, based on the sub-millimeter (JCMT-SCUBA) maps, we
obtained the cold dust mass to be $\sim$ 30 $M_{\odot}$ implying a gas mass of
$1.3\times10^4$ $M_{\odot}$.

From the radio image at 1280 MHz band, we see that this H II region
 consists of seven radio sources, of which S21-S25 form a complex
which is separated from
the two compact sources, S26 and S27. This is consistent with what TFT
obtained in their
high resolution VLA mapping of this source at 4.9 and 8.5 GHz. While
S26 and S27 can be identified with TFT11 and TFT12, respectively, TFT30 is
resolved into S21-S25.
At 610 MHz, these sources are unresolved. The
image at 325 MHz shows a single source at this frequency. The ZAMS spectral
types as obtained from the radio flux densities from our maps and
the maps of TFT are consistent (Table 6).

A comparison of the 1280 MHz radio emission with NIR K$_s$ band
emission is shown in Fig 15.
The two compact radio sources, S26 and S27, do not have any NIR counterparts
which are detected in all the three JHK$_s$ bands. We, however, note that there
are two sources, designated as K1 and K2, detected only in K$_s$-band,
 which match
($<2.5\arcsec$) with S26 and S27, respectively. It is likely that these
objects are deeply embedded. Garay et al. (1986) have used the H$76\alpha$
line and
continuum emission to show that the structure of G45.07+0.13 (positional match
with S27) suggests the existence of an ionised ring. According to them, this
ionised toroidal structure is likely to correspond to the inner ionised walls
of a disk surrounding an inner massive star. This could explain the 2MASS NIR
detection only in K$_s$-band. Since S26 also has compact radio emission and a
NIR counterpart detected only in K$_s$ band, it is likely that
it is also in a similar stage of evolution. Kraemer et al. (2003) have
studied this source as a part of their Galactic Ring Survey in the mid infrared
 (12.5 \& 20.6 $\micron$) and they find three mid infrared compact sources,
one located at the position of S26 and two close (2$\arcsec$) compact sources
near S27; the brighter of them matches in position with S27. The OH maser
position (Argon et al., 2000) and CO bipolar outflow detected by Hunter et al.
(1997) is at the position of S27. Hanson et al. (2002) have reported that this
source is heavily reddened in the K band spectrum ($A_K=8$) and no emission is
seen in the H band spectrum.

Considering the radio nebulosity encompassing S21-S25, we find that
there are four NIR sources within it. IR8 is very bright in J (8.1) as well
as K$_s$ (6.8) bands. Although lying much higher than the ZAMS curves in the
CM diagram, it does not have an associated maser or outflows (as in the case
of IR4). It is, therefore, likely to be a foreground source rather than an
embedded early type star associated with this source. We attribute the radio
emission to another source, designated K3, detected only in H and
K$_s$ bands. K3 is of
ZAMS spectral type O6.5-O6 from the (H-K vs K) CM diagram. The other two
sources within the radio nebulosity are of later spectral types (later than B1).
Since K3 does not match in position with any of the radio sources,
we explore whether S21, ... , S25 are clumps of dense gas externally excited
by K3. For this, the role of geometric dilution is considered as discussed for
S13 in Section 5.1. We find that only S22 and S24 can barely be excited by
K3, while S21, S23 and S25 cannot be ionised by K3.

In this HII region, we find that there are 2 MSX PSC sources,
G045.0668+00.1372 and G045.0711+00.1325 (designated M3 \& M4,
respectively; listed in Table 9 and shown in
Fig 15), within the radio nebulosity. M4 is close to K1 and K2 while M3
can be associated with K3. As in the case of IRAS 19111+1048, we obtain the
2MASS-MSX colours ($F_{21}/F_8$ and $F_8/F_K$, Lumsden et al. 2002) and
observe that M3 and M4 lie in the general region covered by compact H II
regions.

\subsection{Comparison of properties of IRAS 19111+1048 and IRAS 19110+1045}

The properties of the two neighbouring star forming regions as obtained from
infrared and radio studies have been compared here. The physical
extent of ionised gas around IRAS 19111+1048 (2.8 pc) is larger than its
neighbour, IRAS 19110+1045 (1.3 pc). A comparison
of optical depth of the cold dust, $\tau_{100}$, as obtained from the
radiative transfer model indicates that the dust around IRAS
19111+1048 is optically thin ($\tau_{100}\sim0.002$) compared to that
around IRAS 19110+1045 ($\tau_{100}\sim0.1$). The ratio of total
luminosity to the cloud mass of IRAS
19111+1048 and IRAS 19110+1045 are
170 $L_{\odot}/M_{\odot}$ and $24$ $L_{\odot}/M_{\odot}$, respectively.
 The lower value of $L/M$ of IRAS 19110+1045 indicates
that it is younger than its neighbouring star forming region. The near
infrared data as well as high resolution radio
measurements indicate that IRAS 19111+1048 has a larger number of ZAMS stars
energising compact and evolved H II regions.
On the other hand, IRAS 19110+1045 shows the
presence of seven radio sources.
The NIR counterparts of two of the compact H II sources (S26 and S27)
have been
detected only in the K$_s$ band indicating them to be
deeply embedded in dust.
In view of the above, viz,
 the lower $L/M$ ratio of IRAS 19110+1045,
 its smaller extent of ionised
gas, fewer ZAMS stars, the presence of
compact H II regions harbouring deeply embedded stars, IRAS
19110+1045 is likely to be younger than IRAS 19111+1048
as suggested by others (Hunter et al. 1997; Kraemer et al. 2003).

\section{Summary}

The star forming regions associated with IRAS 19111+1048
and IRAS 19110+1045 have been studied at
infrared (near, mid and far), sub-millimeter as well as
low frequency radio wavebands. The dust
environment of these regions is probed
using data from the TIFR balloon-borne telescope
(130 \& 200 $\micron$), MSX, IRAS-HIRES, JCMT-SCUBA and ISO. Using
2MASS PSC, the NIR sources in these regions have been
extracted and studied using colour-magnitude (CM)
diagrams. The distribution of ionised gas has been
obtained using high angular resolution radio maps
from GMRT (325, 610 \& 1280 MHz). The radio morphology indicates presence
of a highly inhomogeneous medium. Twenty compact radio
sources have been extracted from IRAS 19111+1048 and seven from IRAS
19110+1045. One of the compact sources (S20) is found
to be of non-thermal nature based on its
spectral index ($\alpha=-0.6$). The rest of the
sources in IRAS 19111+1048 region are of spectral type earlier than
B0.5.
Based on these sources,
the power law index of the initial mass function has been found to be
$5.3\pm0.5$ for the mass range
$14<m/M_{\odot}<33$.
The compact source, S14,
is located at the position of this IRAS source. The near infrared (NIR) source coincident with the IRAS peak is likely to be an embedded pre-main
sequence star due to its large infrared excess as well as the presence of OH
maser and CO bipolar outflow here. The
electron temperature and emission
measure of S14 using results from GMRT (our work) and VLA (TFT)
measurements are
determined to be $8346\pm1511$ K and $8.8\pm0.7\times10^6$
cm$^{-6}$pc, respectively. A comparison of NIR and radio images has been
carried out to identify the exciting sources
and their nature.
The spectral types determined from NIR and radio measurements
are found to match very well for a good fraction of radio sources having
NIR counterparts.
The radio and infrared morphology of IRAS 19110+1045 indicates that the
extent of dust as well ionised gas is relatively smaller than that of
 IRAS 19111+1048.
The two compact radio sources, S26 and S27, have been
detected only in the K$_s$-band, indicating these to be
deeply embedded.
Self-consistent radiative transfer modelling has been carried out through
spherical gas-dust clouds for both the IRAS sources,
using a cluster of OB stars as determined from the radio maps.
A constant radial density distribution ($n(r) \propto r^0$) is preferred.
The geometric details of the gas-dust clouds, the dust composition and
optical depths etc. have been quantified by the models.
A comparative study of
the infrared and radio properties of IRAS 19110+1045 and IRAS 19111+1048
support the younger nature of the former, consistent with earlier
studies.

\acknowledgments

We thank the anonymous referee for comments and suggestions which improved the
paper. It is a pleasure to thank several members of the Infrared Astronomy
Group of TIFR for their support during the balloon flight campaigns. All
members of the
TIFR Balloon Facility, Hyderabad, are thanked for their technical support for
the flight. We thank the staff of the GMRT who have made the
radio observations possible. GMRT is run by the National Centre for Radio
Astrophysics of the Tata Institute of Fundamental Research. We thank IPAC,
Caltech, for providing us the HIRES-processed IRAS products.




\clearpage


\begin{table}
\caption{Details of the radio continuum observations carried out using the
Giant Metrewave Radio Telescope (GMRT), India.}
\vspace{1cm}
\begin{tabular}{|l l l l |} \hline \hline
Frequency band (MHz) & 1280 & 610 & 325 \\
\tableline
Date of Observation & 28 Oct 2001 & 06 Jan 2002 & 30/31 Aug 2002 \\
Primary beam & 21$\arcmin.6$ & 54$\arcmin$ & 1$^{\circ}.8$ \\
Synthesized beam & $7\arcsec.5 \times 3\arcsec.8$ & $8\arcsec.4 \times 6\arcsec.6$ & $14\arcsec.4 \times 11\arcsec.5$ \\
Position angle & -34$^{\circ}.4$ & -73$^{\circ}.8$ & 68$^{\circ}.4$ \\
Map noise (mJy/beam) & 0.8 & 0.7 & 1.3 \\
Continuum bandwidth & 16 MHz & 16 MHz & 16 MHz \\
\tableline
\end{tabular}
\end{table}



\begin{table}
\caption{Flux density details of IRAS 19111+1048 and IRAS 19110+1045.}
\vspace{1cm}
\hspace*{-1cm}
\begin{tabular}{|c |c c| c c c c |c c c c|} \hline \hline
 IRAS PSC & \multicolumn{10}{c|}{Flux Density (Jy) for $\lambda$ ($\micron$)} \\ \hline
 Source & \multicolumn{2}{c|}{TIFR images$^a$} & \multicolumn{4}{c|}{IRAS-HIRES images$^a$} & \multicolumn{4}{c|}{MSX images$^a$} \\
 & \multicolumn{2}{c|}{} & \multicolumn{4}{c|}{IRAS PSC} & \multicolumn{4}{c|}{ } \\ \hline
 & 200 & 130 & 100  & 60 & 25 & 12 & 21.3 & 14.7 & 12.1 & 8.3 \\ \hline
 19111+1048 & 1145 & 2889 & 6054 & 7024 & 1607 &
320 & 1411 & 511 & 316 & 119 \\
 &     &      & 7497 & 5913 & 1395 & 250 &  & & & \\
 19110+1045 & - & 2261 & 4409 & 4311 & 567 & 72.0 & 319 & 125 & 66 & 32 \\
 &     &      & $<7497$ & $<5913$ & 494 & 58 & & & & \\ \hline
\end{tabular}

$^a$ Fluxes obtained by integrating over a circular region of diameter $3\arcmin$ centered on the peak. \\
\end{table}



\begin{table}
\caption{List of extracted sources along with their peak and integrated
flux densities obtained from the radio maps of IRAS 19111+1048.}
\vspace{1cm}
\begin{tabular}{|c c c c c c c|} \hline \hline
Source & RA & Dec & \multicolumn{2}{c}{1280 MHz} & \multicolumn{2}{c|}{610 MHz}  \\
 No.  & (2000) & (2000) & Peak flux & Int flux & Peak flux & Int flux \\
 & h m s & $^{\circ}$ $\arcmin$ $\arcsec$ & mJy/bm & mJy & mJy/bm & mJy \\ \hline
 S1 & 19 13 23.0 & +10 54 20 & $16.5\pm0.5$ & $77.0\pm2.7$ & $-$ & $-$ \\
 S2 & 19 13 24.0 & +10 53 42 & $20.5\pm0.5$ & $117.5\pm3.5$ & $-$ & $-$ \\
 S3 & 19 13 24.7 & +10 54 01 & $22.4\pm0.5$ & $81.5\pm2.4$ & $-$ & $-$ \\
 S4 & 19 13 24.7 & +10 54 19 & $28.0\pm0.5$ & $169.2\pm3.7$ & $19.6\pm0.7$ &
$115.2\pm5.2$ \\
 S5 & 19 13 24.9 & +10 54 41 & $23.5\pm0.5$ & $90.2\pm2.2$ & $-$ & $-$ \\
 S6  & 19 13 24.9 & +10 53 48 & $34.5\pm0.5$ & $325.5\pm5.5$ & $22.5\pm0.6$ &
$313.6\pm7.5$ \\
 S7 & 19 13 25.7 & +10 54 00 & $56.4\pm0.5$ & $256.7\pm2.9$ & $25.6\pm0.7 $ &
$221.8\pm9.2$ \\
 S8 & 19 13 25.8 & +10 54 24 & $67.3\pm0.5$ & $328.1\pm3.0$ & $ - $ & $-$ \\
 S9 & 19 13 26.1 & +10 54 12 & $44.8\pm0.5$ & $247.1\pm3.4$ & $-$ & $-$ \\
 S10 & 19 13 26.1 & +10 54 02 & $37.4\pm0.5$ & $157.3\pm2.7$ & $-$ & $-$ \\
 S11 & 19 13 26.9 & +10 54 29  & $101.3\pm0.5$& $1261.7\pm6.3$ & $-$ & $-$ \\
 S12 & 19 13 26.9 & +10 54 17  & $88.3\pm0.5$ & $669.6\pm4.5$ & $-$ & $-$ \\
 S13 & 19 13 27.5 & +10 54 03 & $21.7\pm0.5$ & $103.7\pm3.0$ & $-$ & $-$ \\
 S14 & 19 13 27.8 & +10 53 36 & $487.8\pm0.5$ & $2156.6\pm2.8$ & $150.7\pm0.8$ & $833.7\pm5.1$ \\
 S15 & 19 13 28.1 & +10 54 18 & $27.9\pm0.5$ & $136.3\pm3.0$ & $-$ & $-$ \\
 S16 & 19 13 28.4 & +10 54 06 & $26.3\pm0.5$ & $97.9\pm2.5$ & $-$ & $-$ \\
 S17 & 19 13 28.5 & +10 53 44 & $18.1\pm0.5$ & $53.7\pm2.1$ & $-$ & $-$ \\
 S18 & 19 13 28.7 & +10 54 13 & $20.0\pm0.5$ & $97.4\pm2.9$ & $30.0\pm0.8$ & $89.3\pm3.0$ \\
 S19 & 19 13 28.8 & +10 53 55 & $72.4\pm0.5$ & $178.7\pm2.0$ & $43.5\pm0.8$ &
 $154.7\pm3.5$\\
 S20$^a$ & 19 13 29.7 & +10 53 58 & $20.0\pm0.5$ & $21.1\pm 1.0$ &
$29.4\pm0.9$ & $30.6\pm1.6$ \\ \hline
\end{tabular}

$^a$ Non-thermal source \\
\end{table}


\begin{table}
\caption{List of extracted sources along with their peak and integrated
flux densities obtained from the radio map of IRAS 19110+1045.}
\vspace{1cm}
\begin{tabular}{|c c c c c c c|} \hline \hline
Source & RA & Dec & \multicolumn{2}{c}{1280 MHz} & \multicolumn{2}{c|}{610 MHz}
 \\
 No.  & (2000) & (2000) & Peak flux & Int flux & Peak flux & Int flux \\
 & h m s & $^{\circ}$ $\arcmin$ $\arcsec$ & mJy/bm & mJy & mJy/bm & mJy \\ \hline
 S21 & 19 13 19.4 & 10 50 51 & $27.1\pm0.6$ & $124.1\pm2.2$ & $-$ & $426.1^a\pm4.0$ \\
 S22 & 19 13 20.1 & 10 50 48 & $45.9\pm0.6$ & $183.3\pm2.2$ & $-$ & $426.1^a\pm4.0$ \\
 S23 & 19 13 20.3 & 10 50 59 & $32.4\pm0.6$ & $87.5\pm1.8$ & $-$ & $426.1^a\pm4.0$ \\
 S24 & 19 13 20.4 & 10 50 49 & $42.9\pm0.4$ & $81.9\pm1.4$ & $-$ & $426.1^a\pm4.0$ \\
 S25 & 19 13 20.8 & 10 50 59 & $23.2\pm0.6$ & $28.0\pm1.0$ & $-$ & $426.1^a\pm4.0$ \\
 S26 & 19 13 21.7 & +10 50 48 & $56.0\pm0.6$ & $69.5\pm1.3$ & $-$ & $34.9^b\pm1.3$ \\
 S27 & 19 13 22.0 & +10 50 52 & $30.6\pm0.6$ & $44.5\pm1.4$ & $-$ & $34.9^b\pm1.3$ \\ \hline
\end{tabular}

$^a$ Same value of integral flux density is given as S21 - S25 are unresolved \\
$^b$ Same value of integral flux density is given as S26 \& S27 are unresolved \\
\end{table}



\begin{table}
\caption{List of radio (GMRT and VLA) and near infrared sources along with
their spectral types around IRAS 19111+1048.}
\vspace*{1cm}
\begin{tabular}{|c |c |c |c |c |c| } \hline \hline
\multicolumn{2}{|c|}{Radio (GMRT)} & \multicolumn{2}{c|}{Radio (VLA)$^a$} & \multicolumn{2}{c|}{NIR (2MASS)}  \\ \hline
 Name & Spec type & Name & Spec type & Name & Spec type \\ \hline
 S1 & B0-O9.5 & - & - & - & - \\
 S2 & B0-O9.5 & - & - & - & - \\
 S3 & B0-O9.5 & - & - & - & - \\
 S4 & B0-O9.5 & - & - & - & - \\
 S5 & B0-O9.5 & - & - & - & - \\
 S6 & O9 & TFT13 & O9.5 & IR1 & B0-O9 \\
 S7 & O9.5-O9 & - & - & - & - \\
 S8 & O9 & - & - & - & - \\
 S9 & O9.5-O9 & - & - & - & - \\
 S10 & B0-O9.5 & - & - & - & - \\
 S11 & O7 & - & - & IR2 &  O9-O6 \\
 S12 & O8 & - & - & IR3 & B2 \\
 S13 & B0.5-B0 & - & - & - & - \\
 S14 & O6.5-O6 & TFT15 & O6 & IR4, IR5 & X$^b$, O5 \\
 S15 & B0-O9.5 & - & - & - & - \\
 S16 & B0.5-B0 & - & - & IR6 & B7-B3 \\
 S17 & B0.5-B0 & - & - & - & - \\
 S18 & B0-O9.5 & TFT16 & B0-O9.5 & - & - \\
 S19 & B0-O9.5 & TFT17 & B0-O9.5 & IR7 & B7-B3 \\ \hline
\end{tabular}

$^a$ Work of Testi, Felli \& Taylor (1999) \\
$^b$ Either a pre-main sequence star or compact unresolved stars (see Section 5) \\
\end{table}



\begin{table}
\caption{List of radio (GMRT and VLA) and near infrared sources along with
their spectral types around IRAS 19110+1045 region.}
\vspace*{1cm}
\begin{tabular}{|c |c |c |c |c |c| } \hline \hline
\multicolumn{2}{|c|}{Radio (GMRT)} & \multicolumn{2}{c|}{Radio (VLA)$^a$} & \multicolumn{2}{c|}{NIR (2MASS)}  \\ \hline
 Name & Spec type & Name & Spec type & Name & Spec type \\ \hline
 S21 & B0-O9.5 & - & - & - & - \\
 S22$^b$ & B0-O9.5 & - & - & K3 & O6.5-O6 \\
 S23$^b$ & B0-O9.5 & - & - & K3 & O6.5-O6 \\
 S24 & B0-O9.5 & - & - & - & - \\
 S25 & B0.5-B0 & - & - & - & - \\
 S26 & B0-O9.5 & TFT11 & O9.5-O8.5 & K1 & - \\
 S27 & B0.5-B0 & TFT12 & B0-O9.5 & K2 & - \\ \hline
\end{tabular}

$^a$ Work of Testi, Felli \& Taylor (1999) \\
$^b$ Radio knots which could be thermally excited by K3.\\
\end{table}



\begin{table*}
\begin{center}
\caption{Details of 2MASS PSC sources which are likely to be associated with
GMRT sources at 1280 MHz (see Tables 5 and 6) in IRAS 19111+1048 and IRAS
19110+1045 regions.}
\vspace*{1cm}
\hspace*{-1.5cm}
\begin{tabular}{|c c c c c c c |} \hline \hline
2MASS PSC & Name$^a$ & $\alpha_{2000}$ & $\delta_{2000}$ & J & H & K$_s$\\
designation & & (deg) & (deg) & (mag) & (mag) & (mag) \\ \hline
\multicolumn{7}{|c|}{IRAS 19111+1048} \\ \hline
 J19132485+1053474 & IR1 & 288.353577 & 10.896520 & $14.64\pm0.04$ & $13.11\pm0.04$ & $12.35\pm0.05$ \\
 J19132689+1054176 & IR3 & 288.362067 & 10.904891 & $15.23\pm0.13$ & $14.14\pm0.13$
& $12.83\pm0.12$ \\
 J19132692+1054269 & IR2 & 288.362188 & 10.907485 & $13.10\pm0.03$ & $12.03\pm0.04$
& $11.50\pm0.04$ \\
 J19132786+1053364 & IR4 & 288.366123 & 10.893447 & $11.87\pm0.03$ & $9.97\pm0.04$ & $7.85\pm0.03$ \\
 J19132812+1053297 & IR5 & 288.367173 & 10.891603 & $13.76\pm0.06$ & $11.98\pm0.07$
& $10.08\pm0.05$ \\
 J19132823+1054056 & IR6 & 288.367652 & 10.901567 & $16.35\pm0.14$ & $15.21\pm0.09$
& $14.95\pm0.20$ \\
 J19132895+1053560 & IR7 & 288.370642 & 10.898909 & $15.88\pm0.09$ & $14.96\pm0.09$
& $13.58\pm0.10$ \\
\hline
\multicolumn{7}{|c|}{IRAS 19110+1045} \\ \hline
 J19131970+1050478 & IR8 & 288.332099 & 10.846624 & $8.14\pm0.01$ & $7.23\pm0.03$ & $6.83\pm0.02$ \\
 J19132008+1050527 & K3 & 288.333708 & 10.847999 & $-$ & $12.55\pm0.19$ & $11.52\pm0.09$ \\
 J19132184+1050484 & K1 & 288.341025 & 10.846782 & $-$ & $-$ & $13.14\pm0.14$ \\
 J19132208+1050538 & K2 & 288.342034 & 10.848284 & $-$ & $-$ & $10.54\pm0.03$ \\
\hline
\end{tabular}
\end{center}

$^a$ Short name used in the present work (see figures 11, 14 and 15)
\end{table*}



\begin{table*}
\begin{center}
\caption{Best-fit parameters of the radiative transfer models for IRAS
19111+1048 and IRAS 19110+1045.}
\vspace*{1cm}
\hspace*{-1.5cm}
\begin{tabular}{|c c c c c c c c c c c|} \hline \hline
Source & $\alpha$ & $R_{max}$ & $R_{min}$ & $r_{H II}$ & $\tau_{100}$ &
L & Dust & Composition & M$_{dust}$ & Gas:Dust\\
IRAS   &          &   (pc)    &   (pc)    &   (pc)    &              &
(10$^5$ $L_\odot$)  &  Type & Si:C   &  ($M_\odot$) & \\ \hline
19111+1048$^{ }$ & 0.0  & 4.0 & 0.01 & 1.5 &
0.002 &  8.5 & DL & 5:95 & 12 & 250:1\\
19110+1045 & 0.0 & 3.5 & 0.01 & 0.04 & 0.10
 & 3.3 & DL & 60:40 & 450 & 450:1\\ \hline
\end{tabular}
\end{center}
\end{table*}



\begin{table*}
\begin{center}
\caption{Details of the MSX PSC sources within the radio
nebulosity of IRAS 19111+1048 and IRAS 19110+1045 regions.}
\vspace*{1cm}
\hspace*{-1.5cm}
\begin{tabular}{|c c c c c c c c|} \hline \hline
MSX PSC & Name$^a$ & $\alpha_{2000}$ & $\delta_{2000}$ & $F_{8}$ & $F_{12}$ &
$F_{14}$ & $F_{21}$\\
designation & & (deg) & (deg) & (Jy) & (Jy) & (Jy) & (Jy) \\ \hline
 G045.1331+00.1436 & M1 & 288.3610 &  10.9083 & $4.7\pm0.2$ &  $8.3\pm0.9$  & $32.1\pm2.0$ & $57.6\pm3.5$ \\
 G045.1221+00.1323 & M2 & 288.3660 &  10.8933 & $87.5\pm3.6$ & $296.4\pm14.8$ & $455.7\pm27.8$ &  $1061.7\pm63.7$
\\
 G045.0711+00.1325 & M3 & 288.3417 &  10.8482 & $20.3\pm0.8$ & $53.8\pm2.7$ & $119.3\pm7.3$ & $256.0\pm15.4$ \\
 G045.0668+00.1372 & M4 & 288.3354 &  10.8466 & $3.6\pm0.1$ &  $9.0\pm0.5$ & $11.0\pm0.7$ & $35.8\pm2.2$  \\ \hline
\end{tabular}
\end{center}

$^a$ Short name used in the present work (see figures 14 and 15) \\
\end{table*}


\clearpage
\begin {figure}
\epsscale{1}
\plottwo{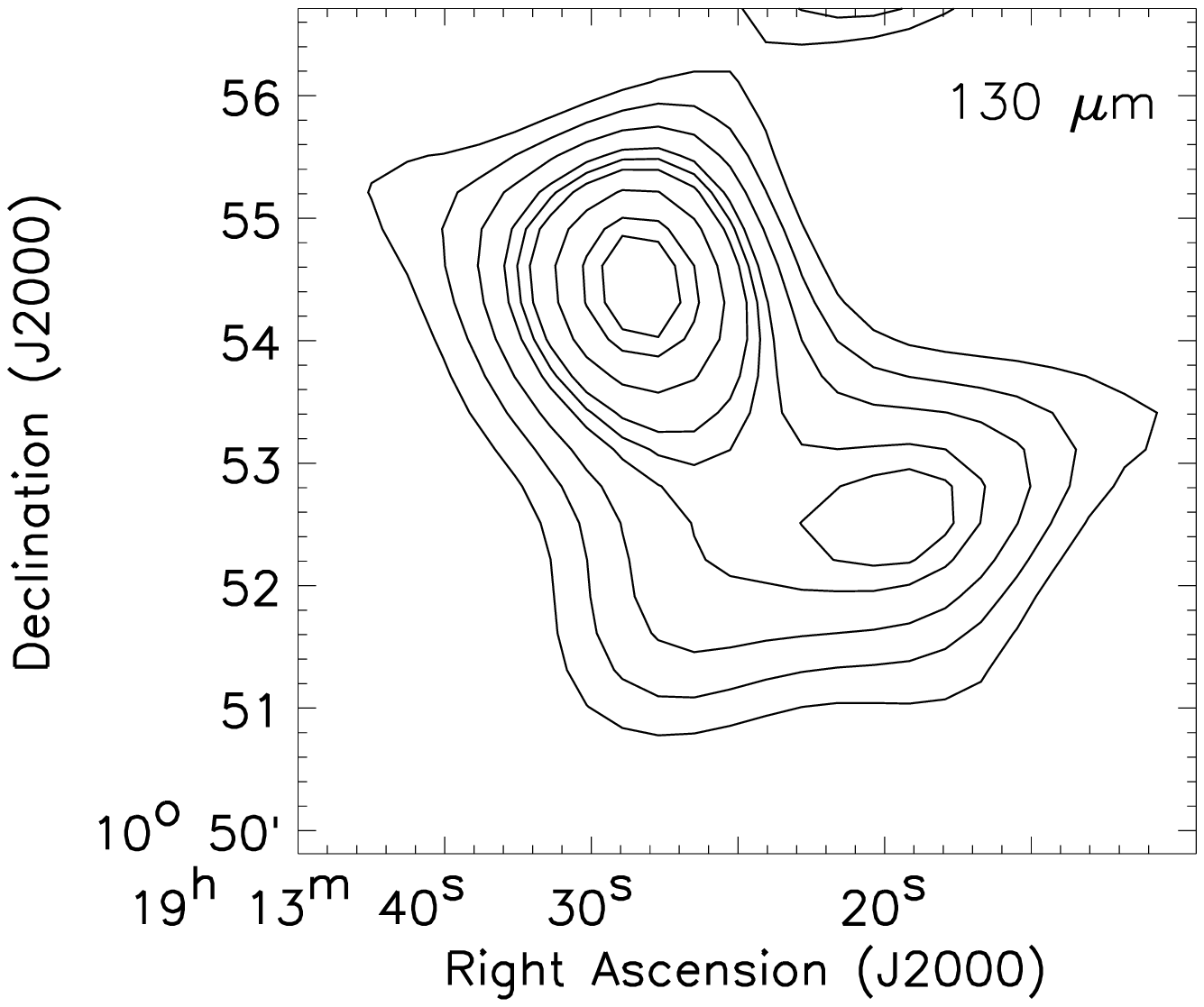}{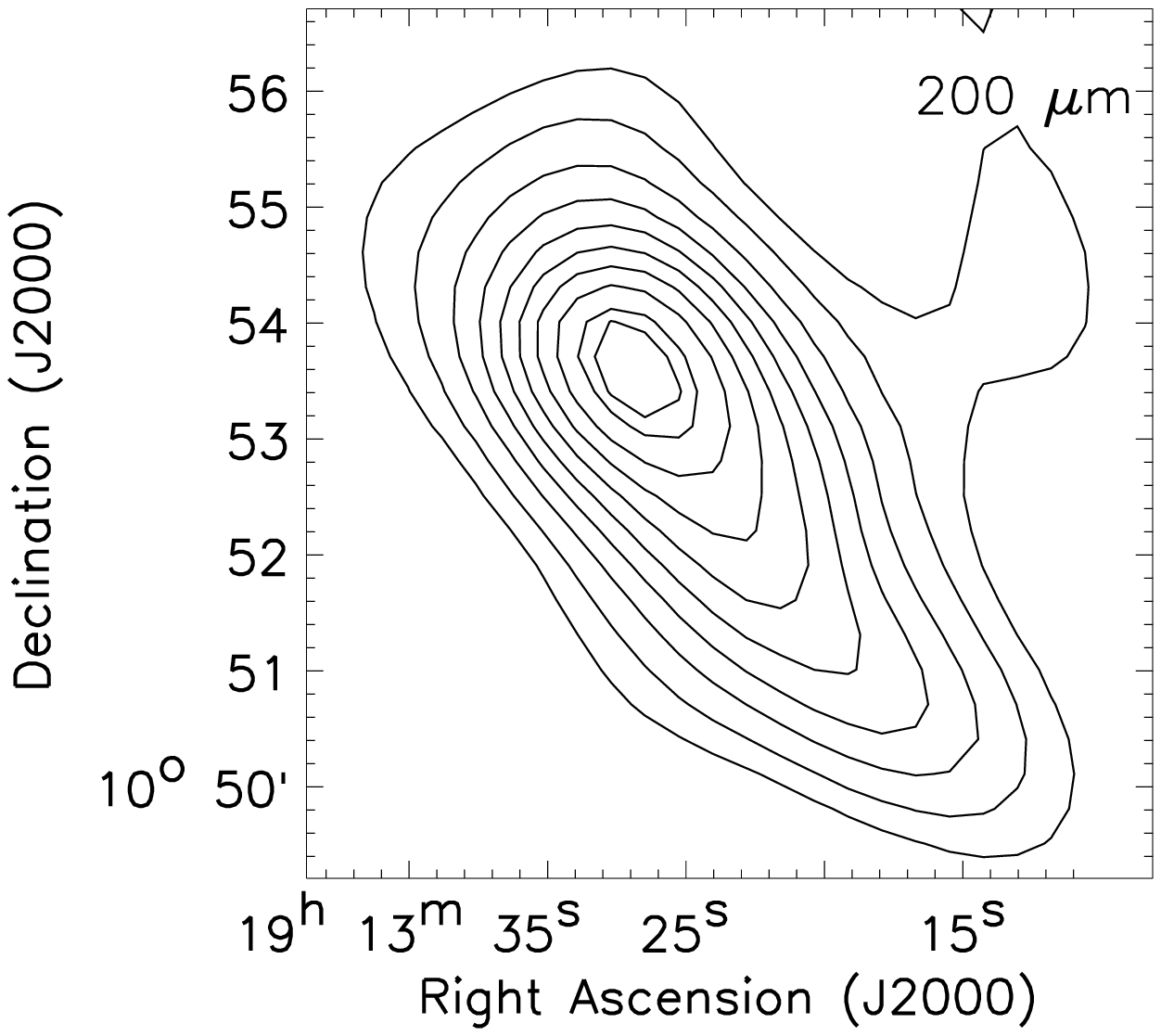}
\caption{The intensity map for the region around IRAS 19111+1048 which
includes IRAS 19110+1045 at 130 $\mu$m (left) and 200 $\mu$m (right).
The effective beamsize for these observations is $\sim1\arcmin.7$. (Left)
Contour levels are at 30, 40, 50, 60, 65, 70, 80, 90,
95\% of peak intensity of 620 Jy arcmin$^{-2}$. (Right) Contour levels are at
12, 20, 30, 40, 50, 60, 70, 80, 90, 95\% of peak intensity of 261 Jy arcmin$^{-2}$.
\label{fig1}}
\end {figure}

\begin {figure}
\epsscale{1}
\plottwo{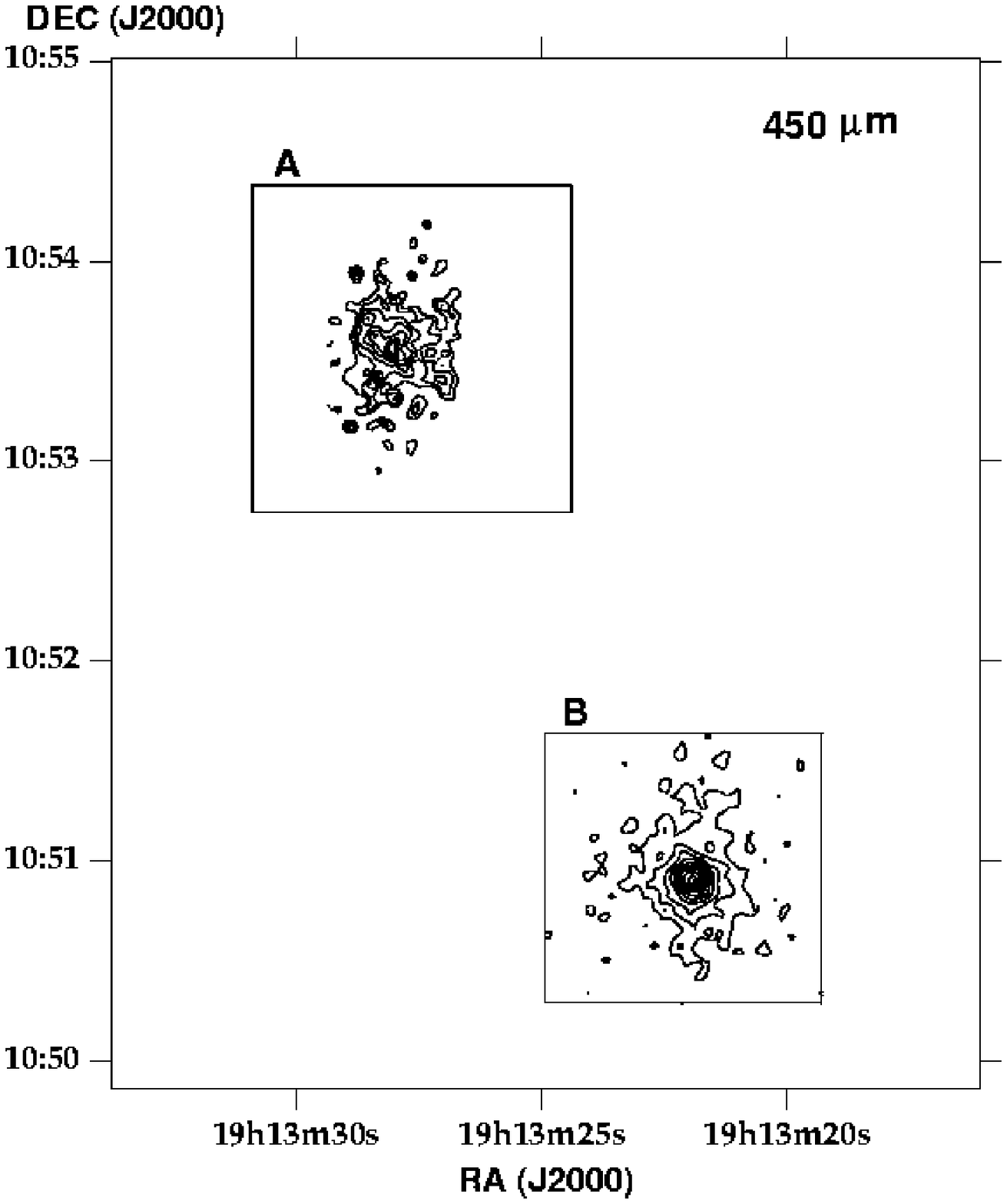}{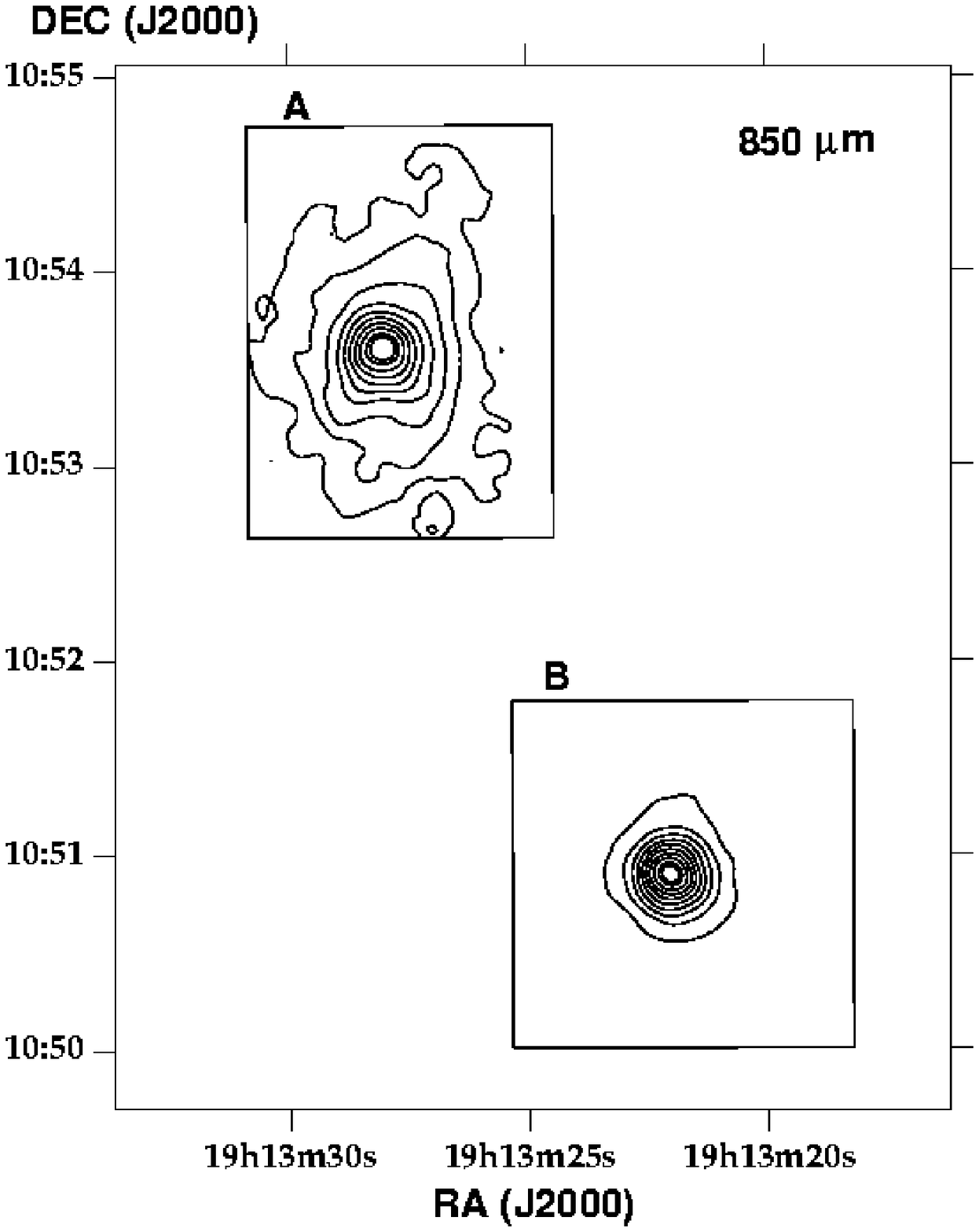}
\caption{Spatial distribution of cold dust emission as obtained from JCMT at
 450 $\micron$ (left) and 850 $\micron$ (right) around the region
IRAS 19111+1048 (A) and IRAS 19110+1045 (B). The contour levels of A are at
40, 50, 60, 70, 80, 90, 95 \% of peak value of 25.5 Jy/beam at 450 $\micron$
and 10, 20, 30, 40, 50, 60, 70, 80, 90, 95 \% of the peak value of 6.9 Jy/beam
at 850 $\micron$. For B, the contour levels are at 10, 20, 30, 40, 50, 60, 70,
80, 90, 95 \% of the peak values of 161.4 Jy/beam and 11.0 Jy/beam at 450 \&
850 $\micron$, respectively. The maps at 450 $\micron$ (left) have a beam size
of $\sim7\arcsec\times11\arcsec$ and at 850 $\micron$ (right) have a beam size
of $\sim13\arcsec\times17\arcsec$.
\label{fig5}}
\end {figure}

\begin {figure}
\plottwo{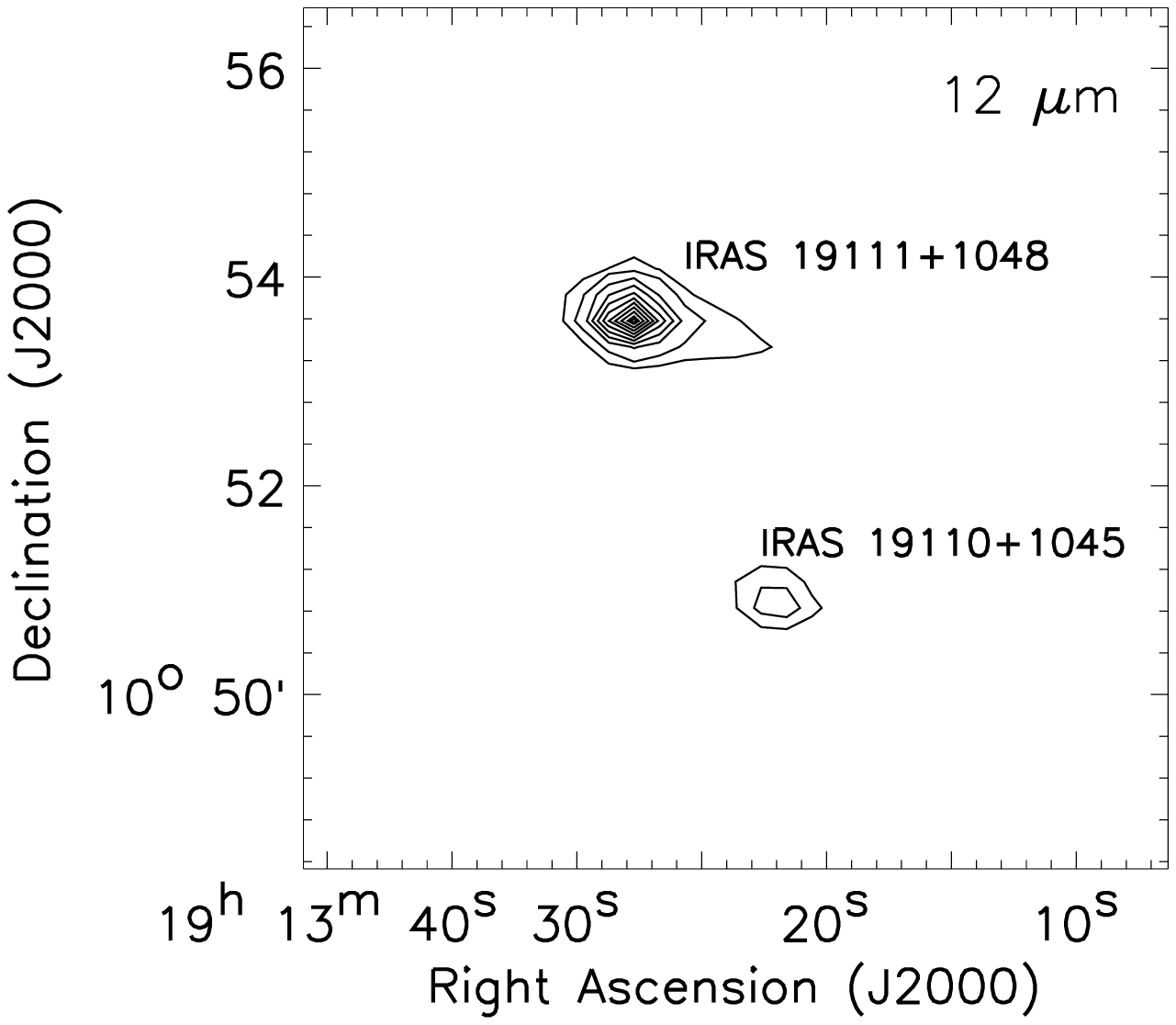}{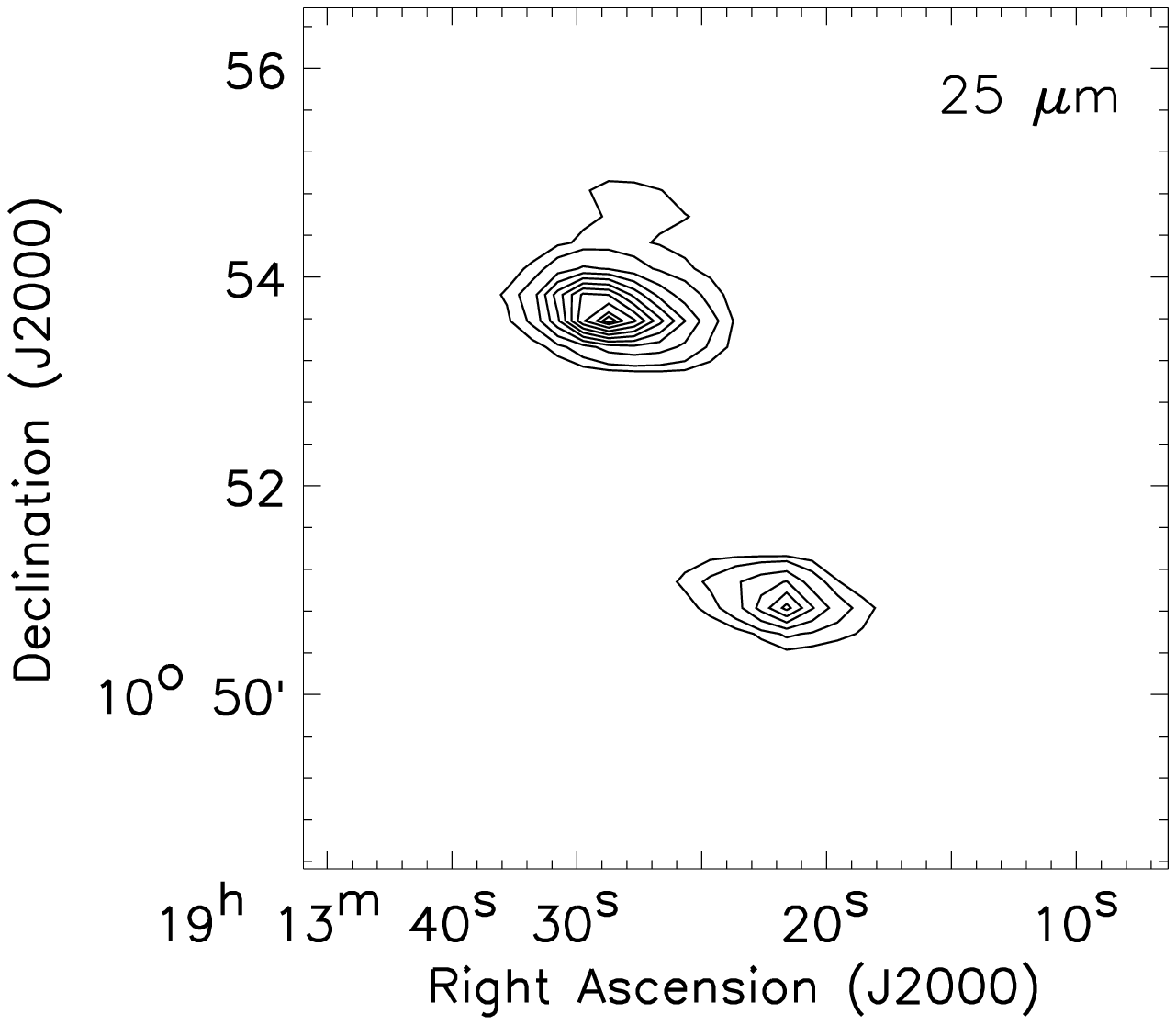}
\plottwo{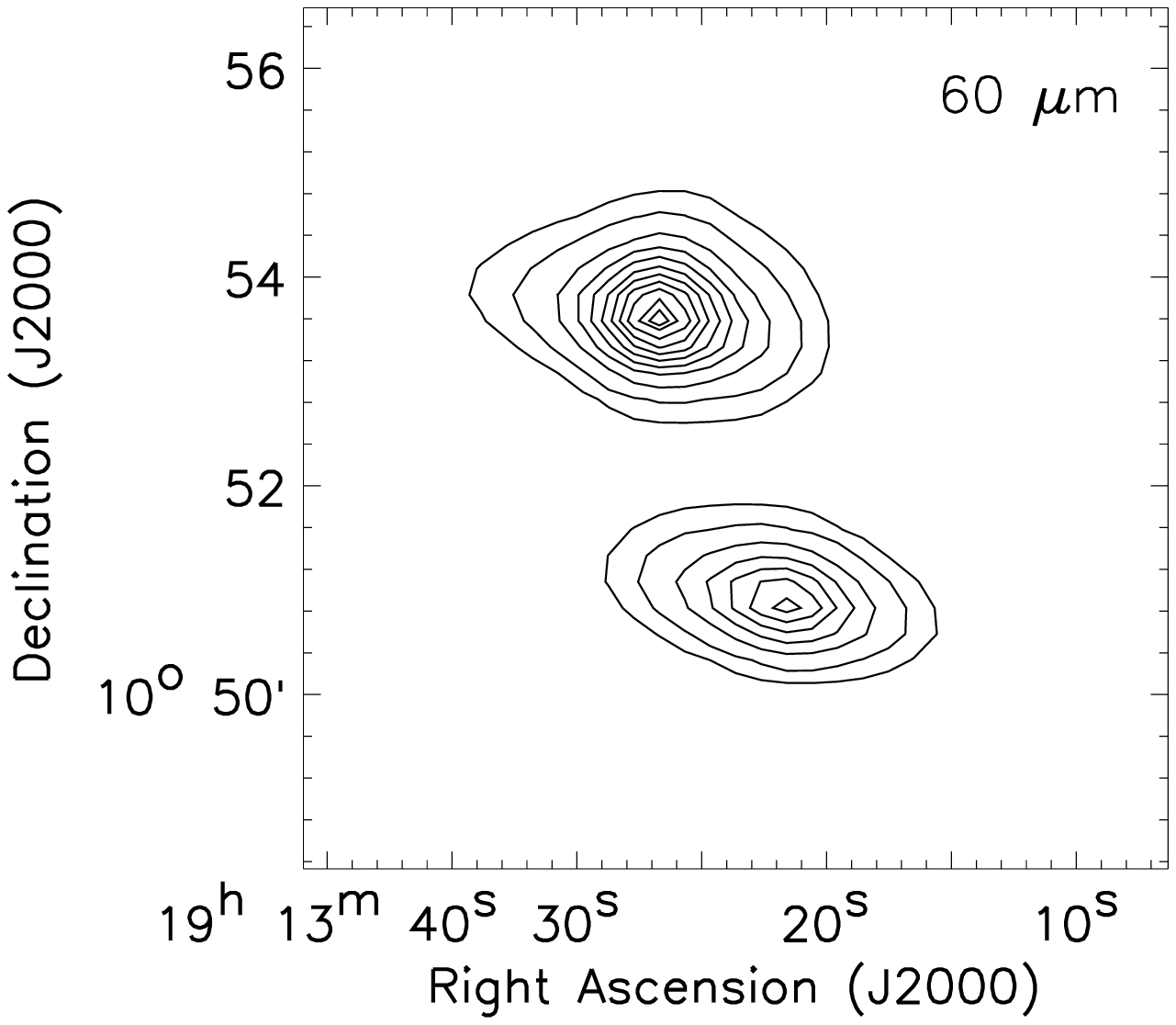}{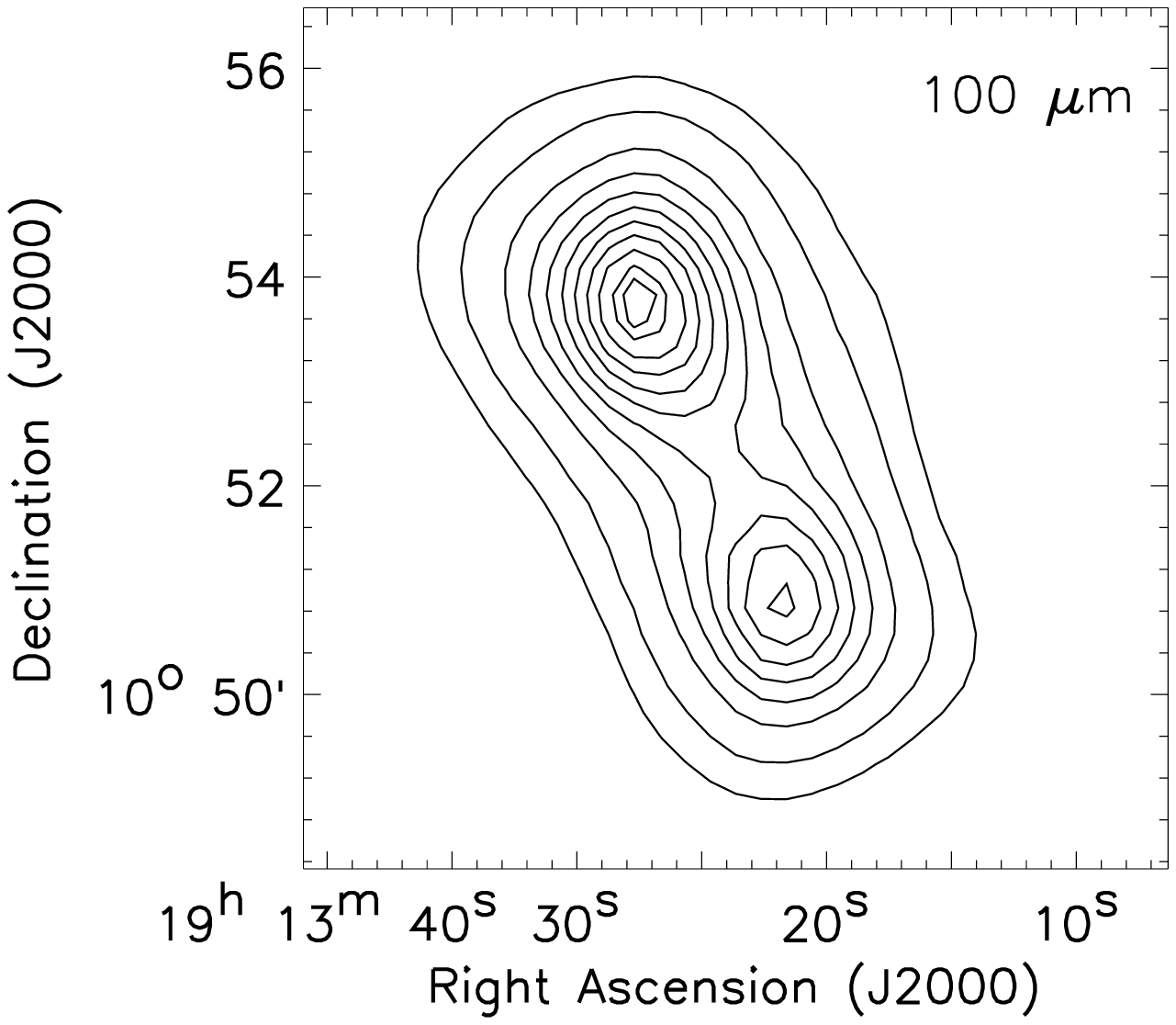}
\caption{The IRAS-HIRES intensity maps for the region covering IRAS 19111+1048
and IRAS 19110+1045 at 12 $\micron$ (top left), 25 $\micron$ (top right), 60
$\micron$ (bottom left) and 100 $\micron$ (bottom right).
The size of the beams are $0\arcmin.75\times4\arcmin.5$ at 12 \micron,
$0\arcmin.75\times4\arcmin.6$ at 25\micron, $1\arcmin.5\times4\arcmin.7$
 at 60 $\micron$ and $3\arcmin.0\times5\arcmin.0$ at 100 $\micron$. However,
the achieved angular resolution of these HIRES maps are much higher (see
text). The contours are at
5, 10, 20, 30, 40, 50, 60, 70, 80, 90 and 95 \% of the peak value of
939 Jy arcmin$^{-2}$, 2350 Jy arcmin$^{-2}$, 5100 Jy arcmin$^{-2}$ and 1970 Jy
arcmin$^{-2}$ at 12, 25, 60 and 100 $\micron$, respectively.
\label{fig3}}
\end {figure}

\begin {figure}
\epsscale{1}
\plotone{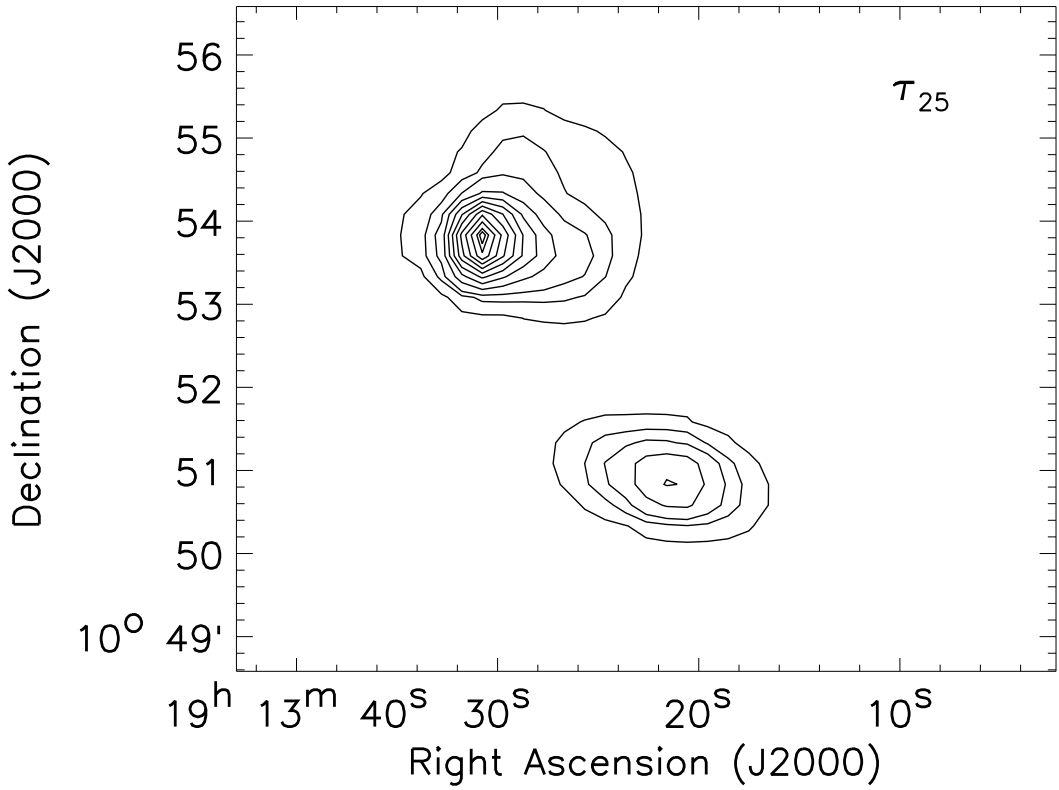}
\caption{The distribution of optical depth at 25 $\mu$m, $\tau_{25}$, of the
region around IRAS 19111+1048 and IRAS 19110+1045 assuming a dust emissivity
law of $\epsilon_{\lambda} \propto \lambda^{-1}$. The $\tau_{25}$ contours
represent  1, 5, 10,  20, 30, 40, 50, 60, 70, 80 and 90 \% of the peak value
of $0.001$.
\label{fig4}}
\end {figure}

\begin {figure}
\epsscale{1}
\plottwo{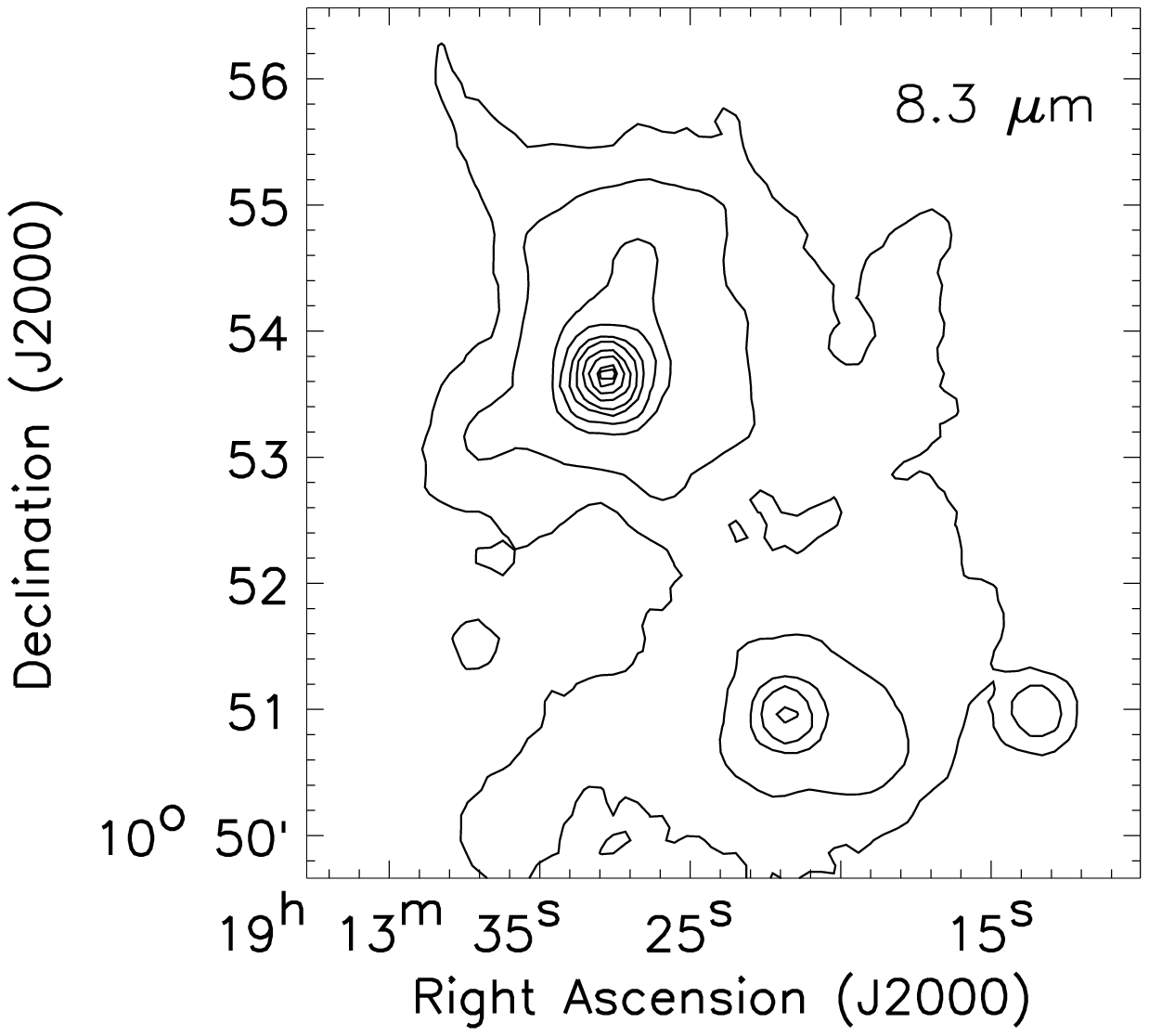}{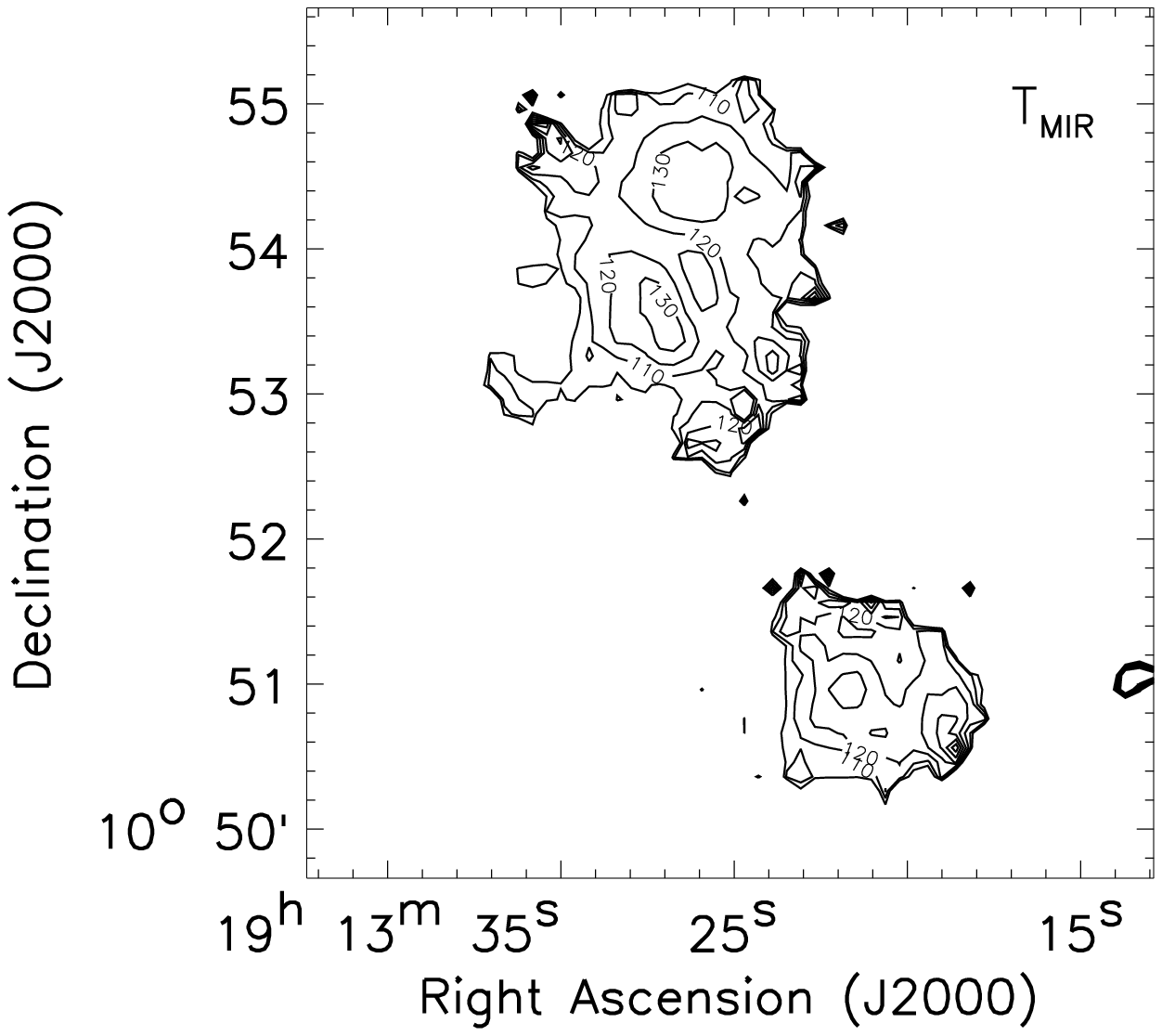}
\caption{The spatial distribution of continuum emission in the MSX 8.3
$\micron$ band (left) with beam $\sim18''$ and mid-infrared
temperature, T$_{\textrm{MIR}}$ (right). The contour levels
are at (left) 0.1, 0.5, 1, 5, 10, 20, 30, 40, 50, 60, 70, 80, 90, 95 \% of
peak value of 473.7 Jy arcmin$^{-2}$ and (right) 100, 110, 120, 130, 140, 150
and 160 K.
\label{fig5}}
\end {figure}

\begin {figure}
\epsscale{1}
\plotone{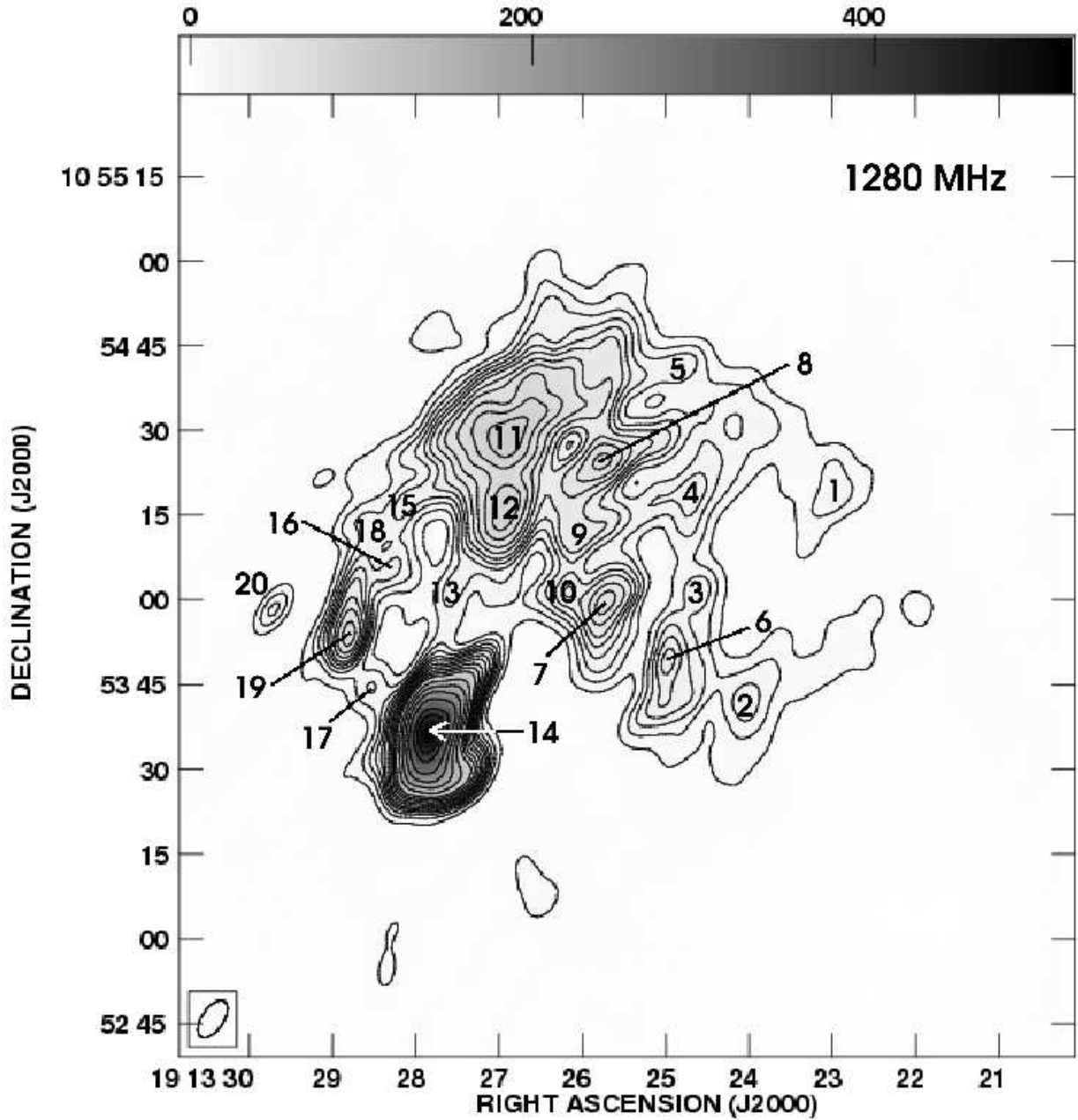}
\caption{Radio continuum emission at 1280 MHz from IRAS 19111+1048. The contour levels are at $3.0\times$(-2, 2, 4, 6, 8, 10, 12, 15, 18, 22, 26, 31, 42, 54, 70, 90, 120, 140) mJy/beam. The beam size is $7\arcsec.5 \times 3\arcsec.8 $ and
rms noise in the map is $\sim 0.8$ mJy/beam. The discrete radio sources are
represented by the numbers as listed in Tables 3 and 5.
\label{fig7}}
\end {figure}

\begin {figure}
\epsscale{1}
\plotone{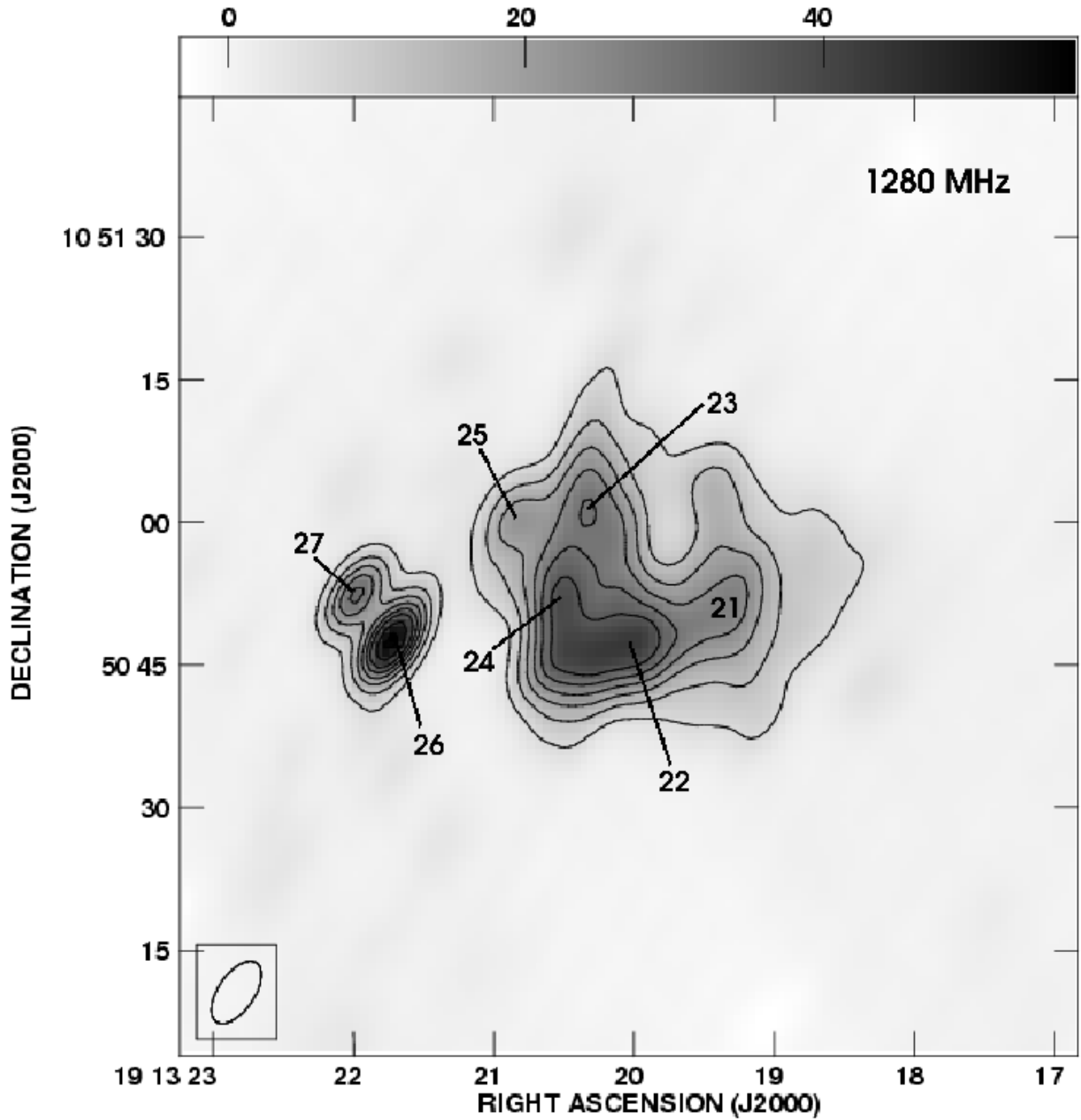}
\caption{Radio continuum emission at 1280 MHz from IRAS 19110+1045. The contours are at levels $3.0\times$(-2, 2, 4, 6, 8, 10, 12, 14, 16, 18) mJy/beam. The
beam size is $7\arcsec.5 \times 3\arcsec.8 $ and rms noise in the map is
$\sim 0.8$ mJy/beam. The discrete radio sources are
represented by the numbers as listed in Tables 4 and 6.
\label{fig7}}
\end {figure}

\begin {figure}
\epsscale{1}
\plottwo{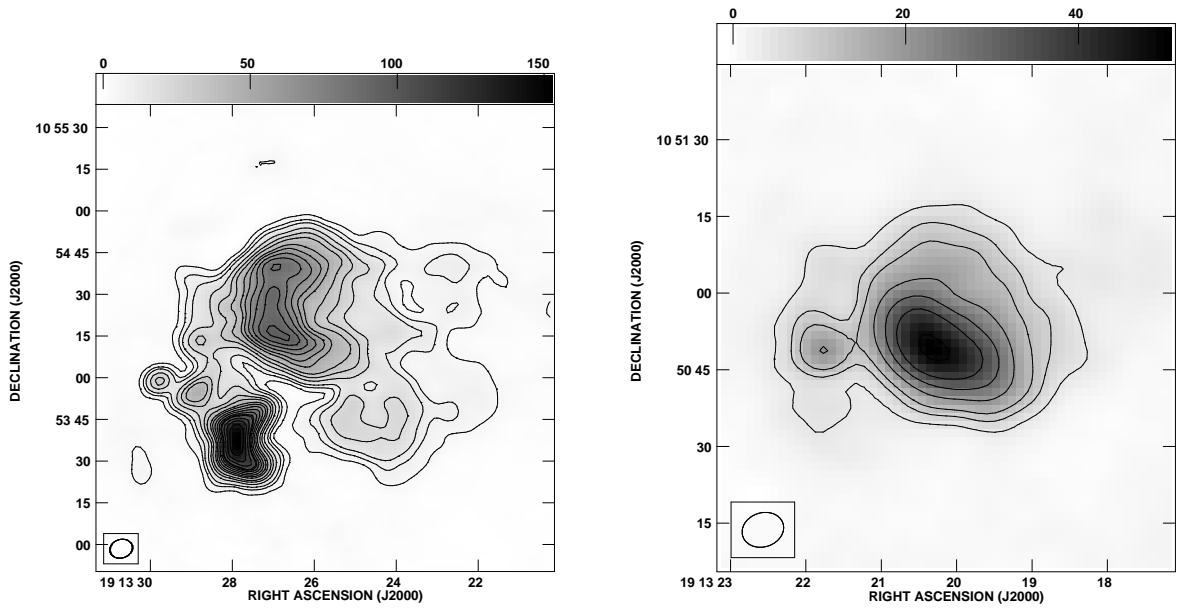}{f8b.ps}
\caption{Radio continuum emission at 610 MHz from IRAS 19111+1048 (left) and
IRAS 19110+1045 (right). The contour levels are at (left) $2.0\times$(-2, 2, 4, 6, 10, 14, 18, 24, 28, 33, 38, 43, 47, 49, 55, 61, 68, 75, 82) mJy/beam
and (right) $2.0\times$(-2, 2, 4, 6, 10, 14, 18, 24) mJy/beam. The beam size
is $8\arcsec.2 \times 6\arcsec.6 $ and the rms noise in the map is $\sim 0.7$
mJy/beam.
\label{fig8}}
\end {figure}

\begin {figure}
\epsscale{1}
\plottwo{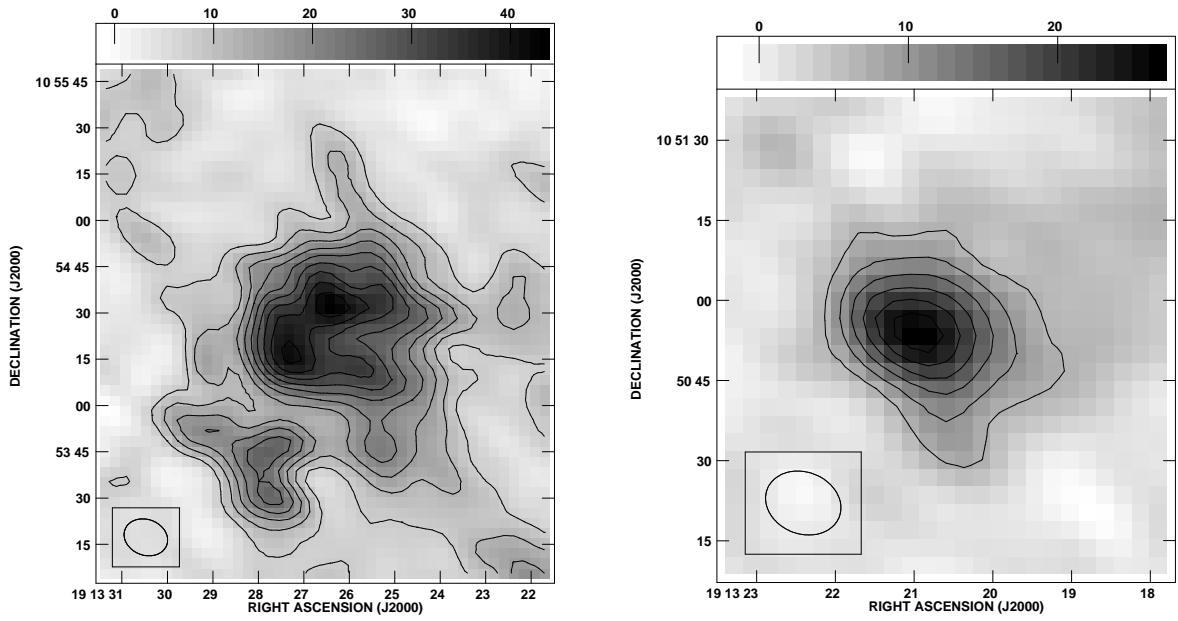}{f9b.ps}
\caption{Radio continuum emission at 325 MHz from IRAS 19111+1048 (left) and
IRAS 19110+1045 (right). The contour levels are at (left) $4.0\times$(-2, 2, 3,
4, 5, 6, 7, 8, 9, 10) mJy/beam and (right) $3.5\times$(-2, 2, 3, 4, 5, 6, 7)
mJy/beam. The beam size is $14\arcsec.5 \times 11\arcsec.4 $ and
map noise is $\sim1.3$ mJy/beam.
\label{fig9}}
\end {figure}

\begin {figure}
\epsscale{0.5}
\plotone{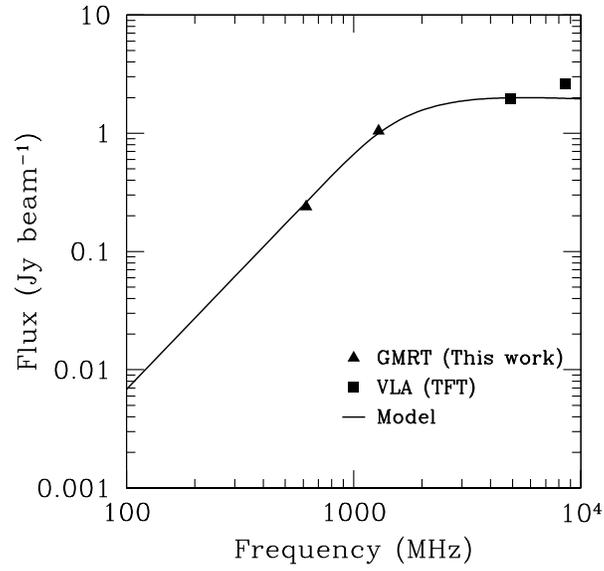}
\caption{The observed (triangles and squares represent GMRT \&
VLA measurements, respectively) and modelled (line) flux densities
for the core of the H II region, S14, in IRAS 19111+1048.
The flux densities correspond to a beam size of
$10\arcsec$.
\label{fig10}}
\end {figure}

\begin {figure}
\epsscale{1}
\plottwo{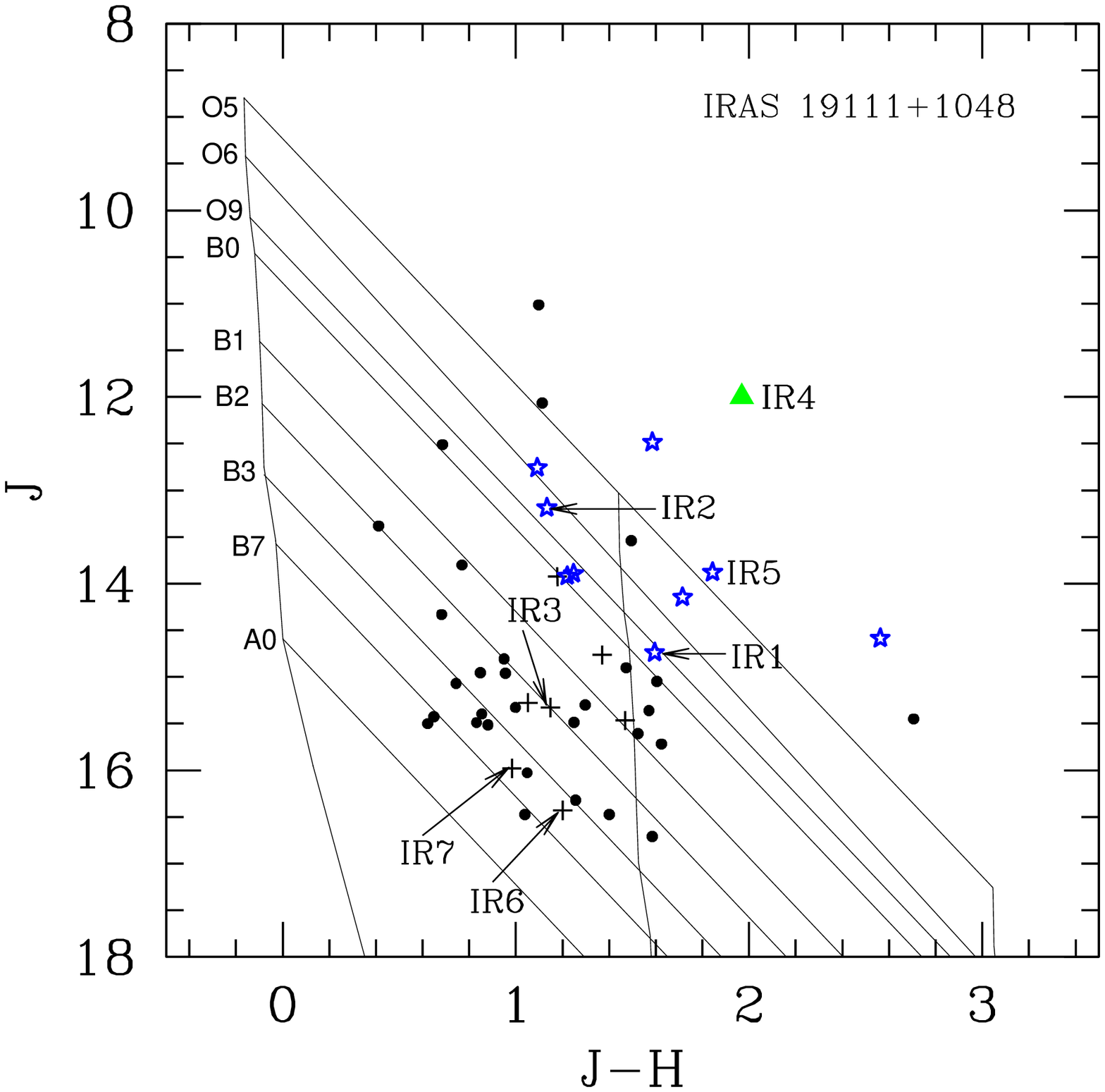}{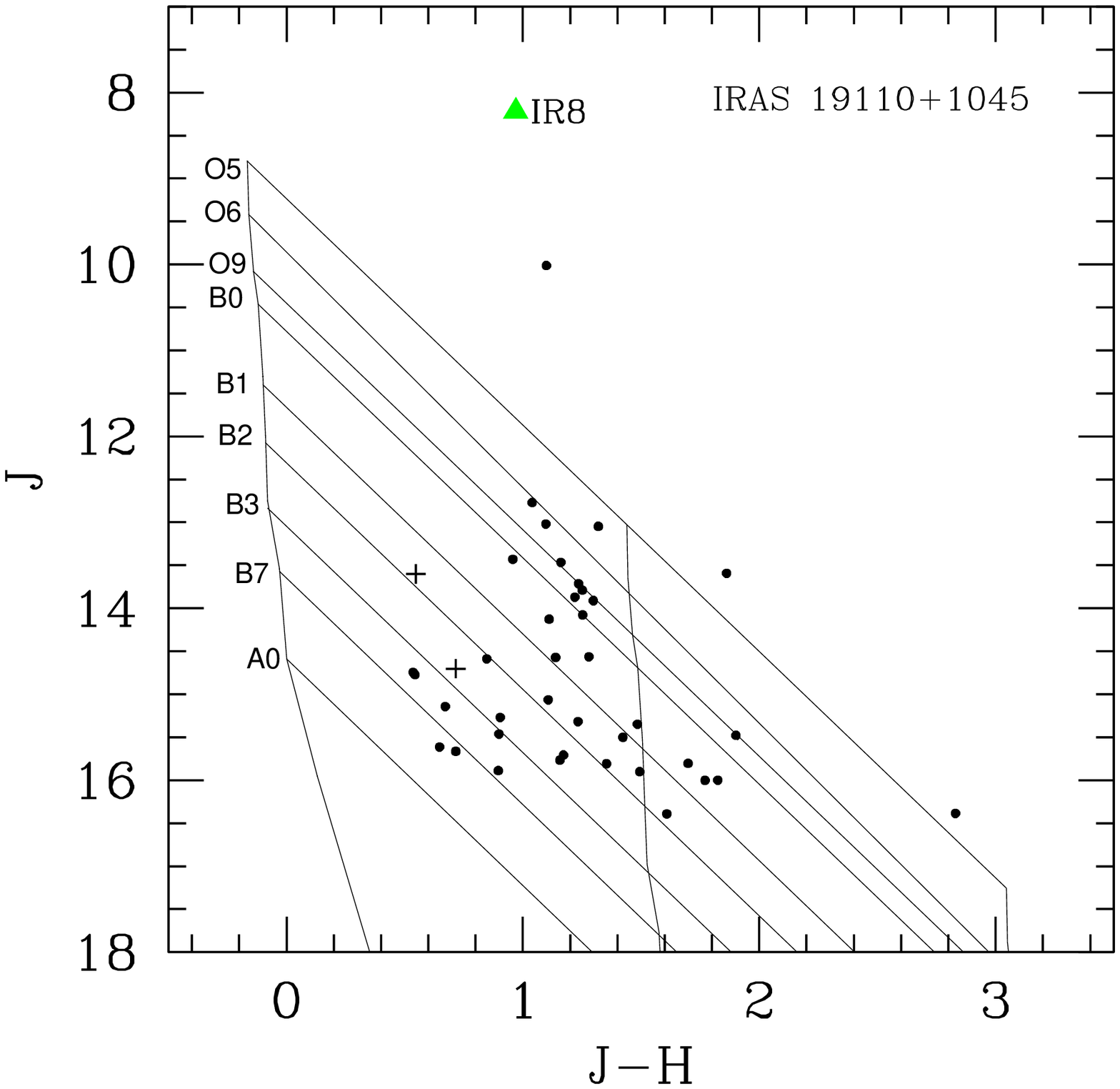}
\caption{Color-magnitude diagram (J-H vs J) for sources detected in all the
2MASS three bands for region around IRAS 19111+1048 (left) and IRAS 19110+1045
(right).
The nearly vertical solid lines from left to right represent the zero
age main sequence (ZAMS) curves reddened by A$_V$ = 0, 15 and 30 mag,
respectively. The slanting lines trace the reddening vectors of these ZAMS
stars.
The asterisk symbols represent stars of spectral type B0 or
earlier. The triangle represents an embedded source, possibly in its very early
evolutionary stage or unresolved early type stars. The plus symbols are
sources of spectral type later than B0 within the radio nebulosity. The dot
symbols represent selected sources not lying within the radio nebulosity.
\label{fig11}}
\end {figure}

\begin {figure}
\epsscale{1}
\plotone{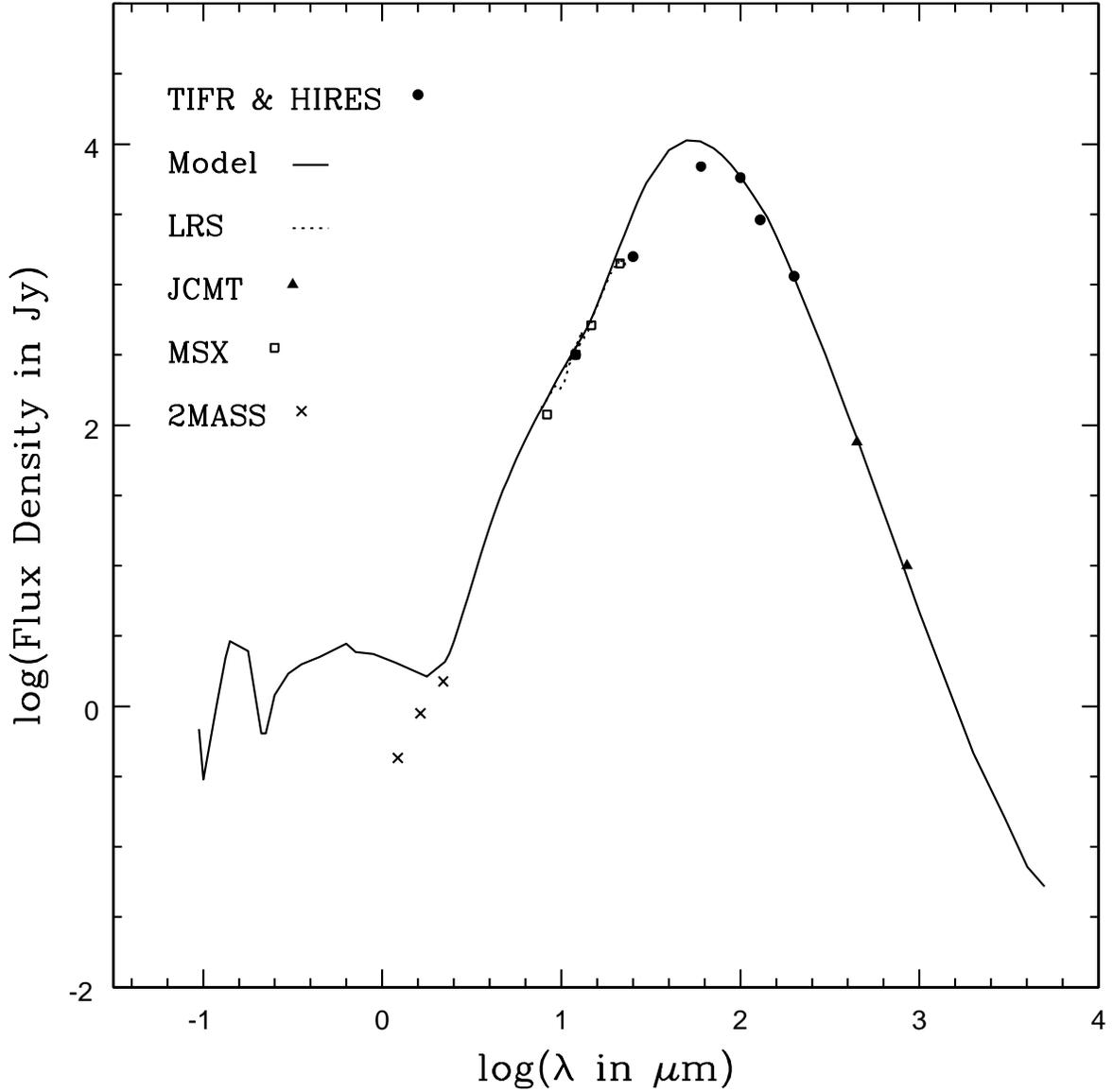}
\caption{Comparison of the spectral energy distribution from
observations and the best fit radiative transfer model of IRAS 19111+1048. The
filled circles represent the TIFR and IRAS-HIRES data. The dashed lines are
the IRAS-LRS spectra. The filled triangles denote JCMT data while the open
squares represent the MSX data. The 2MASS data is denoted by cross-symbols. The
solid line denotes the best fit model to the data.  See text and Table 8 for
details of model parameters.
\label{fig12}}
\end {figure}

\clearpage
\begin {figure}
\epsscale{1}
\plotone{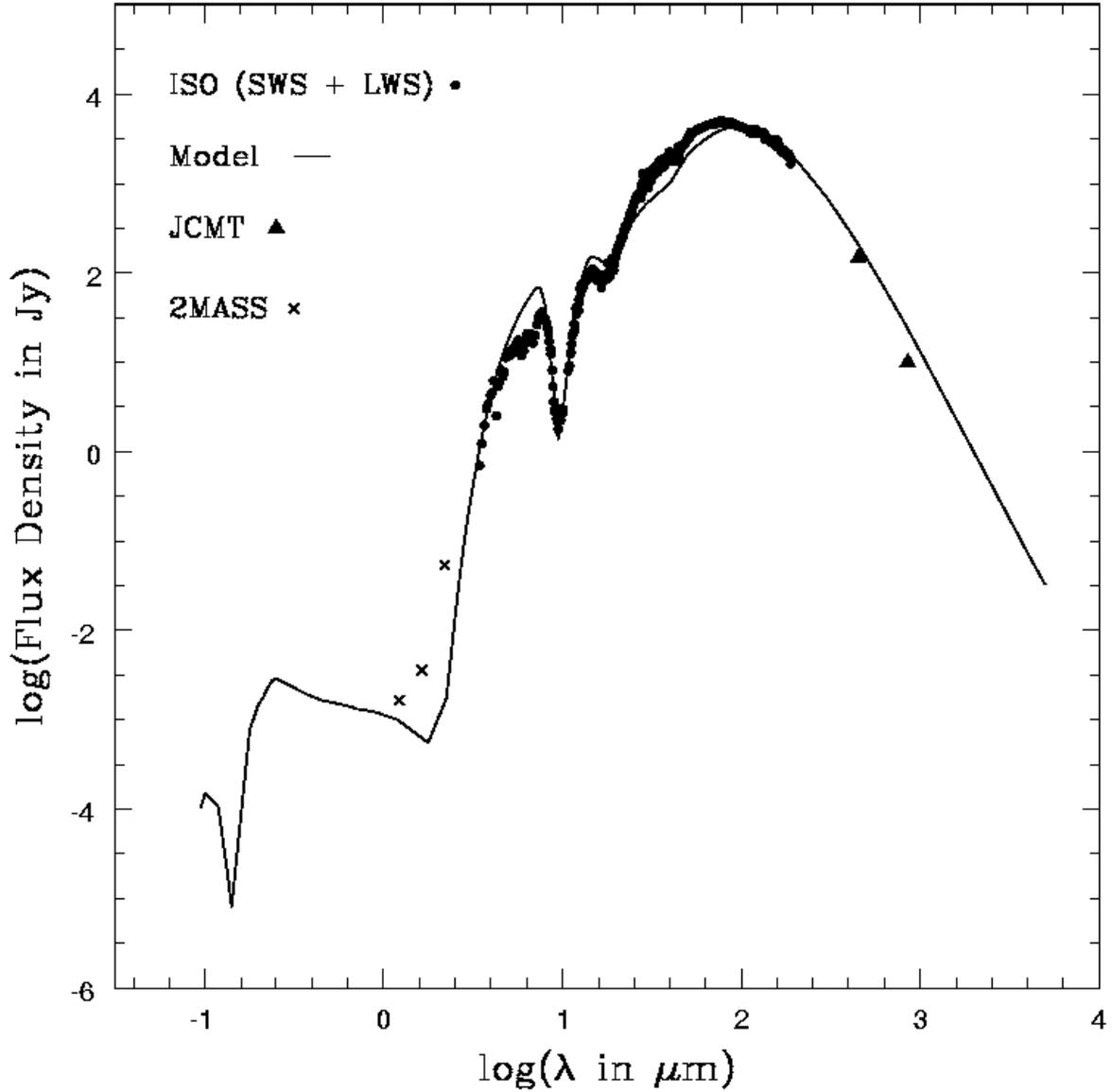}
\caption{Comparison of the spectral energy distribution from
observations and the best fit radiative transfer model of IRAS 19110+1045 for
ISO data. The filled circles represent the ISO SWS and LWS data, the
filled triangles represent JCMT data and the crosses represent the 2MASS data.
The solid line denotes the best fit model to the data. See Table 8 for
details of model parameters.
\label{fig13}}
\end {figure}

\newpage
\begin {figure}
\epsscale{1}
\vspace* {-4cm}
\plotone{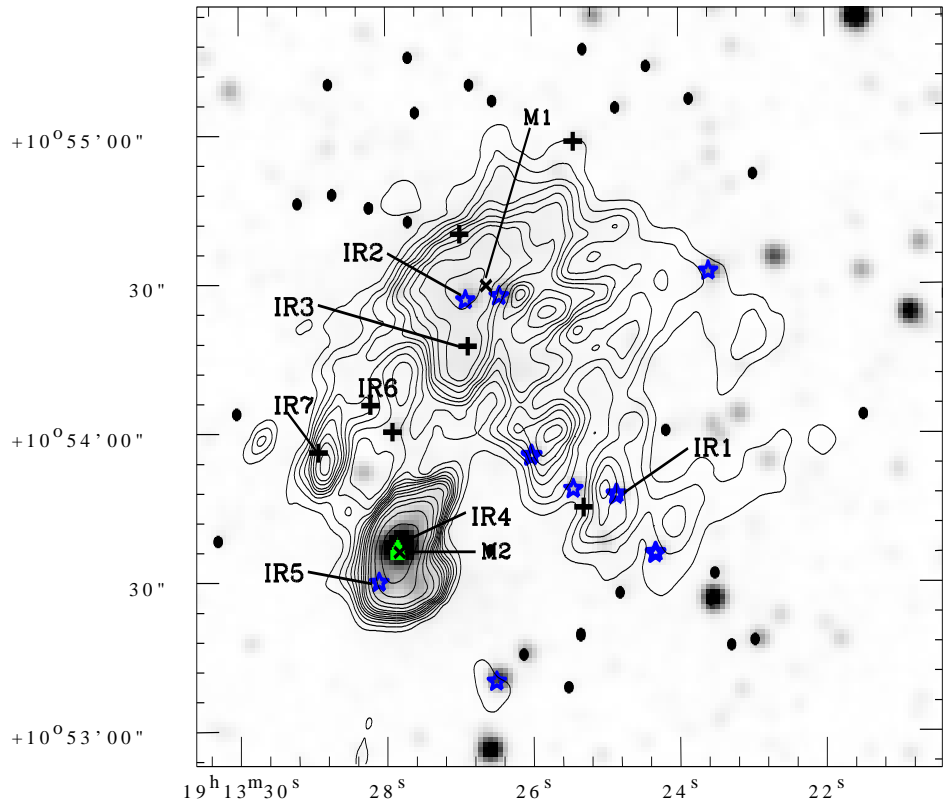}
\vspace* {-5cm}
\caption{Contours of high resolution radio continuum emission at 1280
MHz overlaid over grayscale 2MASS K$_s$-band image for the region
around IRAS 19111+1048. The labelled
axes are in J2000 coordinates. Various NIR and MSX sources are marked. The
cross-symbols represent MSX sources. The description of the other symbols is
the same as given in Fig 11.
\label{fig14}}
\end {figure}

\newpage
\begin {figure}
\epsscale{1}
\vspace* {-4cm}
\plotone{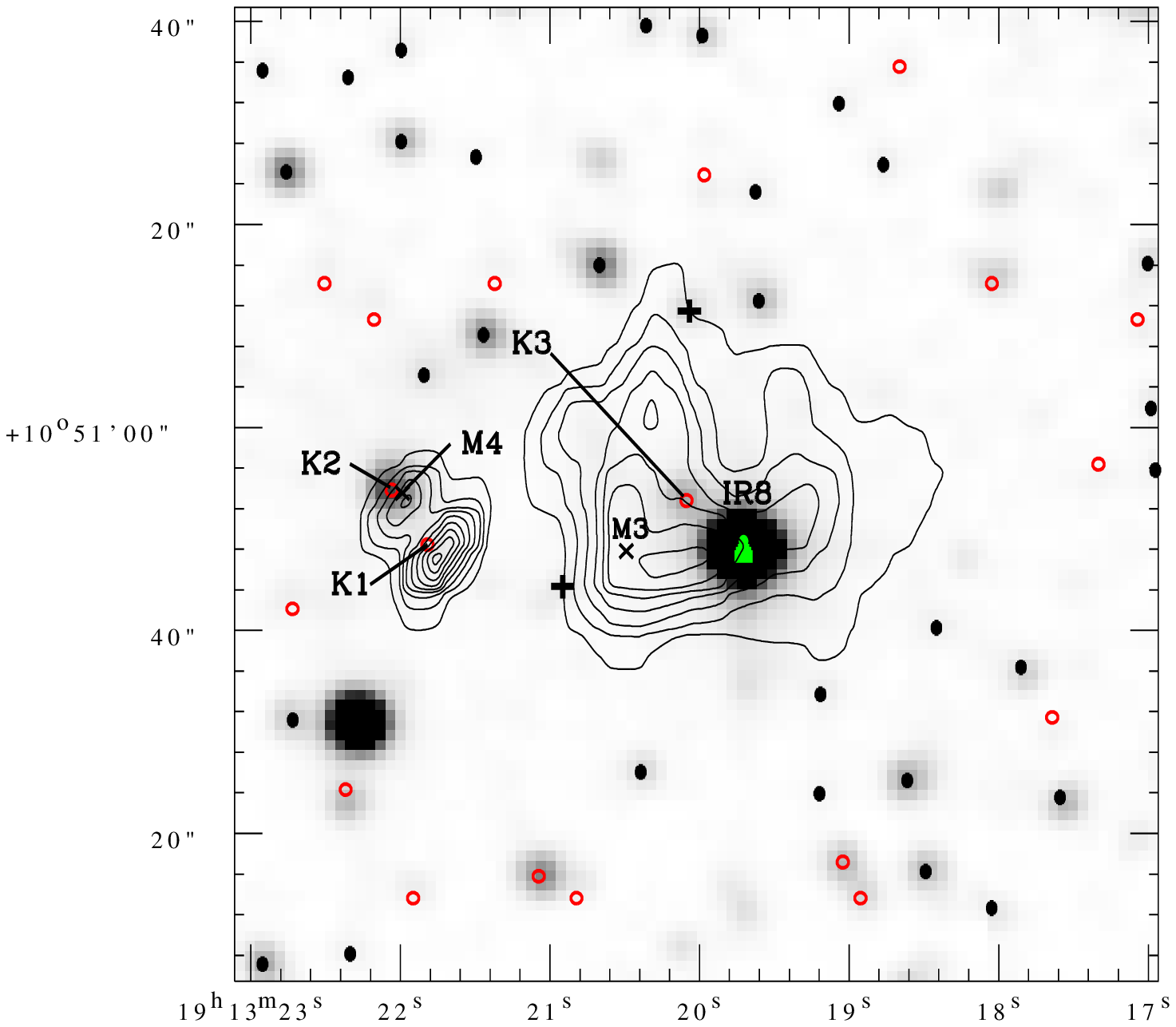}
\vspace* {-5cm}
\caption{Same as in Fig 14 but for the region around IRAS 19110+1045. The
open circles indicate the sources detected only in H and K$_s$ bands
or only in K$_s$-band.
\label{fig15}}
\end {figure}



\begin{thebibliography}{}

\bibitem[1979]{Al79} Altenhoff, W. J., Downes, D., Pauls, T. \& Schraml, J.
1979, A\&AS, 35, 23
\bibitem[2002]{Ar02} Araya, E., Hofner, P., Churchwell, E. \& Kurtz, S. 2002,
ApJS, 138, 63
\bibitem[2000]{Ar00} Argon, A. L., Reid, M. J. \& Menten, K. M. 2000, ApJS, 129, 159
\bibitem[1990]{Au90} Aumann, H. H., Fowler, J. W. \& Melnyk, M. 1990, AJ, 99,
1674
\bibitem[1988]{Be88} Bessel, M. S. \& Brett, J. M. 1988, PASP, 100, 1134
\bibitem[1996]{Cl96} Clegg, P. E., Ade, P. A. R., Armund, C. et al. 1996, A\&A,
 315, L38
\bibitem[2000]{Da00} Darbon, S., Zavagno, A., Perrin, J. -M, Savine, C., Ducci,
 V. \& Sivan, J., -P. 2000, A\&A, 364, 723
\bibitem[1996]{de96} de Graauw, Th., Haser, L. N., Beintema, D. A, et al. 1996,
 A\&A, 315, L49
\bibitem[1984]{Dr84} Draine, B. T. \& Lee, H. M. 1984, ApJ, 285, 89
\bibitem[2003]{Eg03} Egan, M. P., Price, S. D., Kraemer, et al.
2003, Air Force Research Laboratory Technical Report, AFRL-VS-TR-2003-1589
\bibitem[1998]{Fa98} Faison, M., Churchwell, E., Hofner, P., Hackwell, J., Lynch, D. K. \& Russell, R. W. 1998, ApJ, 500, 280
\bibitem[2003]{Fi03} Fish, V. L., Reid, M. J., Wilner, D. J. \& Churchwell, E. 2003, ApJ, 587, 701
\bibitem[2004]{Fr04} Frieswijk, W. F., Shipman, R. F. \& Lahuis, F. 2004, ISO Technical Note 12 
\bibitem[1986]{Ga86} Garay, G., Rodriguez, L. F. \& van Gorkom, J. H. 1986,
ApJ, 309, 553
\bibitem[1977]{Ge77} Genzel, R. \& Downes, D. 1977, A\&AS, 30, 145
\bibitem[1988]{Gh88} Ghosh, S. K., Iyengar, K. V. K., Rengarajan, S. N., Tandon, S. N., Verma, R. P. \& Daniel, R. R. 1988, ApJ, 330, 928
\bibitem[1989]{Gh89} Ghosh, S. K., Iyengar, K. V. K., Rengarajan, T. N., Tandon, S. N., Verma, R. P., Daniel, R. R. \& Ho, P. T. P. 1989, ApJ, 347, 338
\bibitem[2002]{Gh02} Ghosh, S. K. \& Ojha, D. K. 2002, A\&A, 288, 326
\bibitem[1973]{Go73} Goss, W. M., Lockhart, I. A., Fomalont, E. B. \& Hardebeck, E. G. 1973, ApJ, 183, 843
\bibitem[1978]{Gu78} Gull, S. F. \& Daniell, G. J. 1978, Nature, 272, 686
\bibitem[2002]{Ha02} Hanson, M. M., Luhman, K. L. \& Rieke, G. H. 2002, ApJS,
138, 35
\bibitem[1982]{Ha82} Haslam, C. G. T., Stoffel, H., Salter, C. J. \&
Wilson, W. E. 1982, A\&AS, 47, 1
\bibitem[1983]{Hi83} Hildebrand, R. H. 1983, QJRAS, 24, 267
\bibitem[1991]{Ho91} Hoare, M. G., Roche, P. F. \& Glencross, W. M. 1991, MNRAS, 251, 584
\bibitem[1997]{Hu97} Hunter, T. R., Phillips, T. G. \& Menten, K. M. 1997,
ApJ, 478, 283
\bibitem[1986]{IR86} IRAS Science Team, 1986, A\&AS, 65, 607
\bibitem[1982]{Is82} Israel, F. P. 1982, ApJ, 255, 475
\bibitem[2003]{Kr03} Kraemer, K. E., Jackson, J. M., Kassis, M., et al. 2003, ApJ, 588, 918
\bibitem[2003]{Lo03} Lloyd, C., Lerate, M. R. \& Grundy, T. W., 2003,
ISO Technical Note 17
\bibitem[1996]{Lu96} Lumsden, S. L. \& Puxley, P. J. 1996, MNRAS, 281, 493
\bibitem[2002]{Lu02} Lumsden, S. L., Hoare, M. G., Oudmaijer, R. D. \&
Richards, D. 2002, MNRAS, 336, 621
\bibitem[1983]{Ma83} Mathis, J. S., Mezger, P. G. \& Panagia, N. 1983, A\&A, 128, 212
\bibitem[1977]{Ma77} Matthews, H. E., Goss, W. M., Winnberg, A. \& Habing, H. J. 1977, A\&A, 61, 261
\bibitem[1967]{Me67} Mezger, P. G. \& Henderson, A. P. 1967, ApJ, 147, 471
\bibitem[1999]{Mo99} Mookerjea, B. \& Ghosh, S. K. 1999, J. Astrophys. Astr.
20, 1
\bibitem[2000]{Na00} Naik, M. V., D'Costa, S. L., Ghosh, S. K., Mookerjea, B.,
Ojha, D. K. \& Verma, R. P. 2000, Publ. Astron. Soc. Pac., 112, 273
\bibitem[2004]{Oj04} Ojha, D. K., Ghosh, S. K., Kulkarni, V. K., Testi, L., Verma, R. P. \& Vig, S. 2004a, A\&A, 415, 1039
\bibitem[2004]{Oj04} Ojha, D. K., Tamura, M., Nakajima, Y., et al. 2004b, ApJ,
 616, 1042
\bibitem[1973]{Pa73} Panagia, N. 1973, 78, 929
\bibitem[2001]{Pa01} Pankonin, V., Churchwell, E., Watson, C. \& Bieging, J. H. 2001, ApJ, 558, 194
\bibitem[2001]{Pr01} Price, S. D., Egan, M. P., Carey, S. J.,
Mizuno, D. R., \& Kuchar, T. A. 2001, AJ, 121, 2819
\bibitem[1985]{Ri85} Rieke, G. H. \& Lebofsky, M. J. 1985, ApJ, 288, 618
\bibitem[2000]{Sa00} Sandell, G. 2000, A\&A, 358, 242
\bibitem[1986]{Sa86} Sanders, D. B., Clemens, D. B., Scoville, N. Z.\& Solomon, P. M. 1986, ApJS, 60, 1
\bibitem[2001]{Si01} Simon, R., Jackson, J. M., Clemens, D. P. \& Bania, T. M. 2001, ApJ, 551, 747 
\bibitem[1969]{Sc69} Schraml, J. \& Mezger, P. G. 1969, ApJ, 156, 269
\bibitem[2005]{Sh05} Shepherd, D. 2005, Astro-ph 0506024
\bibitem[2003]{Sl03} Sloan, G. C., Kraemer, K. E., Price, S. D. \& Shipman, R. F. 2003, ApJS, 147, 379
\bibitem[1991]{Sw91} Swarup, G., Ananthakrishnan, S., Kapahi, V. K., Rao, A. P.,
 Subrahmanya, C. R., \& Kulkarni, V. K., 1991, Current Science, 60, 95
\bibitem[1999]{Te99} Testi, L., Felli, M. \& Taylor, G. B. 1999, A\&AS, 138, 71 (TFT)
\bibitem[1989]{Wo89} Wood, D. O. S. \& Churchwell, E. 1989, 69, 831

\end{thebibliography}
\end{document}